\documentclass[aps,pre,twocolumn,amsmath,amssymb]{revtex4-2}

\usepackage{mathrsfs}
\usepackage{graphicx}
\usepackage{bm}
\usepackage{threeparttable}
\usepackage{amsmath,amssymb}
\usepackage{color}
\usepackage{epstopdf}
\usepackage{comment}
\usepackage[colorlinks=true,linkcolor=blue,anchorcolor=blue, citecolor=blue,urlcolor=blue]{hyperref}

\def\kbT{{k_BT}}
\def\T{{\theta_{T}}}
\def\Tp#1{{\theta_{T}^{#1}}}
\def\h{\Delta t}
\def\n{\lambda}
\def\a{\alpha}
\def\b{\beta}
\def\g{\gamma}
\def\t{\tau}
\def\s{\sigma}
\def\vp{\varphi}

\def\pr{\prime}
\def\ovl#1{\overline{#1}}

\def\mfr{\mathfrak{m}}

\def\bk#1{\left\langle#1\right\rangle}
\def\ave#1{\mathbb{E}\left[ #1 \right]}
\def\bbkk#1{\kern 0.1em \left\langle \kern -0.3em \left\langle \kern 0.1em #1 \kern 0.1em \right\rangle \kern -0.3em \right\rangle \kern 0.1em}
\def\bko#1{\left( #1 \right)_0}
\def\IntV{\int d\bm{v}}
\def\pd#1{\partial_{#1}}
\def\pdo#1#2{\bko{\pd{#1}{#2}}}
\def\kr#1{\delta_{#1}}
\def\ep#1{\varepsilon_{#1}}

\allowdisplaybreaks[4]
\makeatletter
\renewcommand{\paragraph}{
  \@startsection{paragraph}{4}{\z@}
  {-3.25ex \@plus -1ex \@minus -.2ex}
  {1.5ex \@plus .2ex}
  {\normalfont\footnotesize\bfseries\itshape\centering}
}
\makeatother

\newcommand{\subsubsubsection}[1]{\paragraph{#1}}

\newcommand{\eqr}[1]{Eq.~\eqref{#1}}

\begin{document}

\title{Mesoscale simulation model for odd fluids}

\author{Yuxing Jiao}
\email{jiaoyuxing22@mails.ucas.ac.cn}
\affiliation{Beijing National Laboratory for Condensed Matter Physics and Laboratory of Soft Matter Physics, Institute of Physics, Chinese Academy of Sciences, Beijing 100190, China}\affiliation{School of Physical Sciences, University of Chinese Academy of Sciences, Beijing 100049, China}

\author{Mingcheng Yang}
\email{mcyang@iphy.ac.cn}
\affiliation{Beijing National Laboratory for Condensed Matter Physics and Laboratory of Soft Matter Physics, Institute of Physics, Chinese Academy of Sciences, Beijing 100190, China}\affiliation{School of Physical Sciences, University of Chinese Academy of Sciences, Beijing 100049, China}

\begin{abstract}
A fluid, with broken time-reversal symmetry, would exhibit odd transport coefficients, such as odd viscosity, thermal conductivity and diffusion coefficient, which may fundamentally alter the fluid properties and significantly influence the structure and dynamics of immersed objects. Here, we develop an efficient coarse-grained simulation approach for the odd fluid, that captures all essential features of real odd fluids. Based on microscopic kinetic theory, we analytically derive the transport coefficients of the mesoscale odd fluid. We validate this approach by performing both simulations and theoretical calculations to explore the intricate transport phenomena of the odd fluid under various external drivings. Furthermore, we demonstrate that colloidal particles in the odd fluid exhibit unusual dynamic behavior.
\end{abstract}

\maketitle

\section{Introduction}
In common fluids, transport coefficient tensor, that relates fluxes to drivings, is symmetric (even) under time reversal, which is required by Onsager reciprocal relations~\cite{NET}. However, when time-reversal and parity symmetries of a fluid are broken, its transport coefficient tensor is predicted to contain additional antisymmetric part that are odd under time reversal. Launched six decades ago, the research on odd fluids has now evolved into a field of broad interest. The study of odd fluids encompasses a wide spectrum, spanning from classical to quantum systems and from passive to active systems. The prominent examples include the appearance of odd viscosity and thermal conductivity in polyatomic gases under a magnetic field~\cite{PolyGas,PolyGas2},  odd viscosity in electron Hall systems~\cite{Avron1995,Hoyos2012,Bandurin2019,Holder2019}, odd diffusivity in charged particles in a magnetic field~\cite{Bonella2017,Abdoli_2020,
Sharma2022}, and odd viscosity and diffusivity in chiral active fluids~\cite{ActiveColloidal,
OddEffects1,Vitelli_Fluhydro,Banerjee2017-fe,Vitelli_OddIdealGas,Markovich2021,
Hargus2021,Vega_Reyes2022-vg}. The odd transport coefficients can produce transverse fluxes under longitudinal drivings and give rise to exotic phenomena unattainable in common fluids~\cite{OddEffects1,OddEffects2,OddEffects3,OddEffects4,OddEffects5,Ding2024,Reichhardt_2022,Vitelli_Fluhydro,Komura2021}. 

Besides fundamentally changing the properties of the fluid itself, the odd transport coefficients are expected to importantly influence the structure and dynamics of mesoscopic objects immersed in the fluid. However, the study on such odd complex fluids (composed of mesoscopic objects in odd fluids) is rare and nontrivial, because of its many-body character and the complex interplay between diverse interactions. Even a few existing numerical works~\cite{OddEffects2,OddEffects4,Reichhardt_2022,Hargus2021,Yang_2021}, based on molecular-dynamics-type simulations, mainly focus on a single suspending object, due to the computational bottleneck arising from the drastic difference in time and length scales between solvent and solute particles. In contrast, several coarse-grained fluid approaches have been well established to simulate traditional complex fluids during the past few decades, e.g., lattice Boltzmann method~\cite{LBE,Ladd1_1994,Ladd2_1994}, dissipative particle dynamics~\cite{DPD,Hoogerbrugge_1992} and multi-particle collision dynamics (MPC)~\cite{MPC1,MPC2,MPC_MD,Kapral2008,Gompper_2009}. The coarse-grained approaches extremely speed up the simulations of traditional complex fluids and have achieved considerable success. Hence, it is highly desired to develop an efficient coarse-grained model for the odd fluid, that is critical for exploring the emerging odd complex fluids. 

In this paper, by slightly generalizing the widely-used version of MPC, i.e., stochastic rotation dynamics (SRD), we propose a coarse-grained odd fluid model, named chiral stochastic rotation dynamics (CSRD), that has all odd transport coefficients (odd viscosity, thermal conductivity and diffusion coefficient). The CSRD retains all the advantages of the conventional SRD, such as high efficiency, proper description of hydrodynamics, dissipation, thermal fluctuations, and so on. The simplicity of the CSRD enables us to derive analytical expressions for its transport coefficients. Using simulations and continuum theory, we demonstrate that the CSRD approach correctly captures the intriguing and intricate transport behaviors of odd fluids. Moreover, we show that a dilute odd colloidal suspension possesses unusual dynamics.

\section{The CSRD Model}
Following the conventional SRD, the CSRD fluid is described with a large number of point particles of mass $m$ with positions $\bm{r}_i(t)$ and velocities $\bm{v}_i(t)$. For simplicity, we focus on a two-dimensional (2D) CSRD fluid. Its dynamics consists of alternating streaming and collision steps. In the streaming step, each fluid particle (say $i$) moves ballistically for a certain time $\h$,
\begin{equation}\label{strEq}
    \bm{r}_i(t+\h)=\bm{r}_i(t)+\bm{v}_i(t)\h.
\end{equation}
In the collision step, the particles are sorted into cells of a square lattice with a size $l$, and then their velocities relative to the center-of-mass velocity of each cell are rotated around the $z$ axis by an angle $\alpha$
\begin{equation}\label{colEq}
    \bm{v}_i(t+\h)=\bm{v}_{cm,i}+\bm{R}\left( \alpha \right)\cdot(\bm{v}_i(t)-\bm{v}_{cm,i}),
\end{equation}
with $\bm{v}_{cm,i}=\sum_jm\bm{v}_j/\sum_jm$ the center-of-mass velocity of the collision cell occupied by particle $i$ (the summation runs over all particles in this cell) and $\bm{R}\left( \alpha \right)=\cos\a\bm{I}-\sin\a\bm{\varepsilon}$ the rotation matrix ($\bm{I}$ and $\bm{\varepsilon}$ are the rank-2 identity and Levi-Civita tensors, respectively). To ensure Galilean invariance, the lattice's position is randomly shifted for each collision step~\cite{MPC_RS}. For the SRD, the rotational angle is $\alpha=\Omega$, where $\Omega$ is randomly taken as $\omega$ or $-\omega$ with equal probability (independently for each cell); while, for the CSRD, $\alpha=\Omega+\theta$, where an additional unidirectional constant rotation, $\theta$, is implemented. It is just the additional unidirectional rotation that breaks the time-reversal and parity symmetries, thus allowing the presence of odd transport coefficients that are lacking in the conventional SRD. The role of the nonzero $\theta$ is reminiscent of that played by a constant external magnetic field in charged particle systems. The 2D CSRD can be straightforwardly extended to 3D, where the rotation operation contains two parts: rotation around a stochastic axis (as the traditional 3D SRD) and rotation around a fixed axis (say the z axis) to constantly break the symmetries.

\begin{figure}[htbp!]
    \centering
    \includegraphics[keepaspectratio, width=\columnwidth]{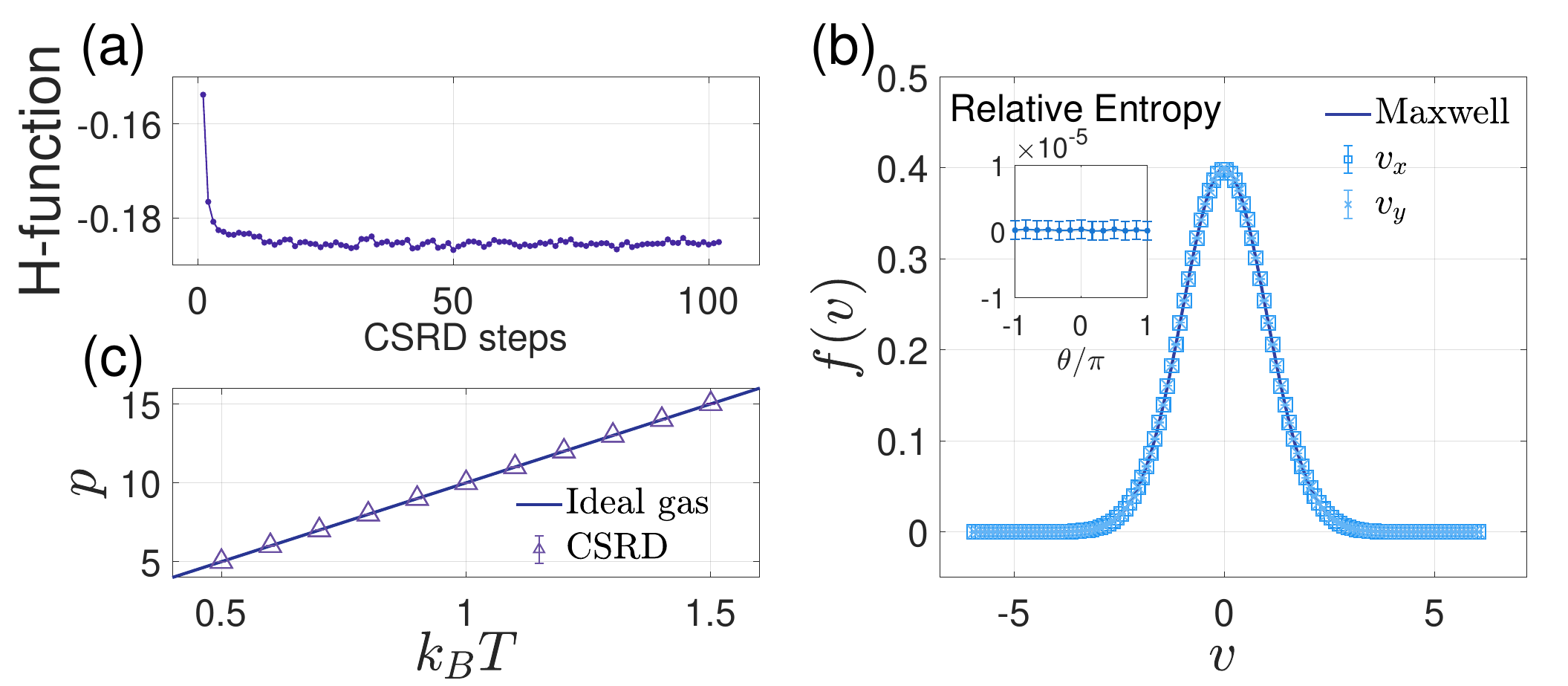}
    \caption{(a) The evolution of H-function in the CSRD simulation. (b) The single-particle velocity distribution, where the symbols and curve correspond to the simulation data and the Maxwellian distribution, respectively. The inset is the vanishing relative entropy at different $\theta$: $D_{KL}\left( f(\bm{v}) || f_{Eq}(\bm{v}) \right)=\int d\bm{v} f(\bm{v})\ln{\left[ f(\bm{v})/f_{Eq}(\bm{v}) \right]}$, which measures the difference between the joint velocity distribution and its equilibrium counterpart. (c) Pressure as a function of temperature with a fixed particle number density $n$. The triangles represent the simulation results for 5 different $\theta$ ($0,\pi/4,\pi/2,3\pi/4,\pi$), which coincide together and agree perfectly with the ideal gas equation of state (solid line). In (a-c), $\omega=2\pi/3$, $n=10$ and $\h=0.1$; and in (a,b) $k_BT=1$ and $\theta=5\pi/9$. In simulations, all quantities are nondimensionalized by setting $m=1$, $l=1$ and the mean thermal energy $=1$.}
    \label{Fig::Equi}
\end{figure}

In addition to allowing the presence of odd transport coefficients, the CSRD fluid retains all other features of the common SRD fluid, which is crucial to properly capture hydrodynamic behavior. For instance, the CSRD locally conserves mass, momentum and energy, and obeys the H-theorem as confirmed in Fig.~\ref{Fig::Equi}(a), in which the H-function, $H(t)=\int d\bm{v} f(\bm{v},t)\ln f(\bm{v},t)$ with $f(\bm{v},t)$ the single-particle velocity distribution, decreases rapidly with time and converges to a steady-state value. Without external perturbation, the CSRD quickly evolves into equilibrium state with a Maxwellian velocity distribution, as shown in Fig.~\ref{Fig::Equi}(b). Moreover, the CSRD fluid satisfies an ideal gas equation of state owing to the absence of potential interaction (Fig.~\ref{Fig::Equi}(c)).

\begin{figure*}[t!]
    \centering
    \includegraphics[keepaspectratio, width=1.8\columnwidth]{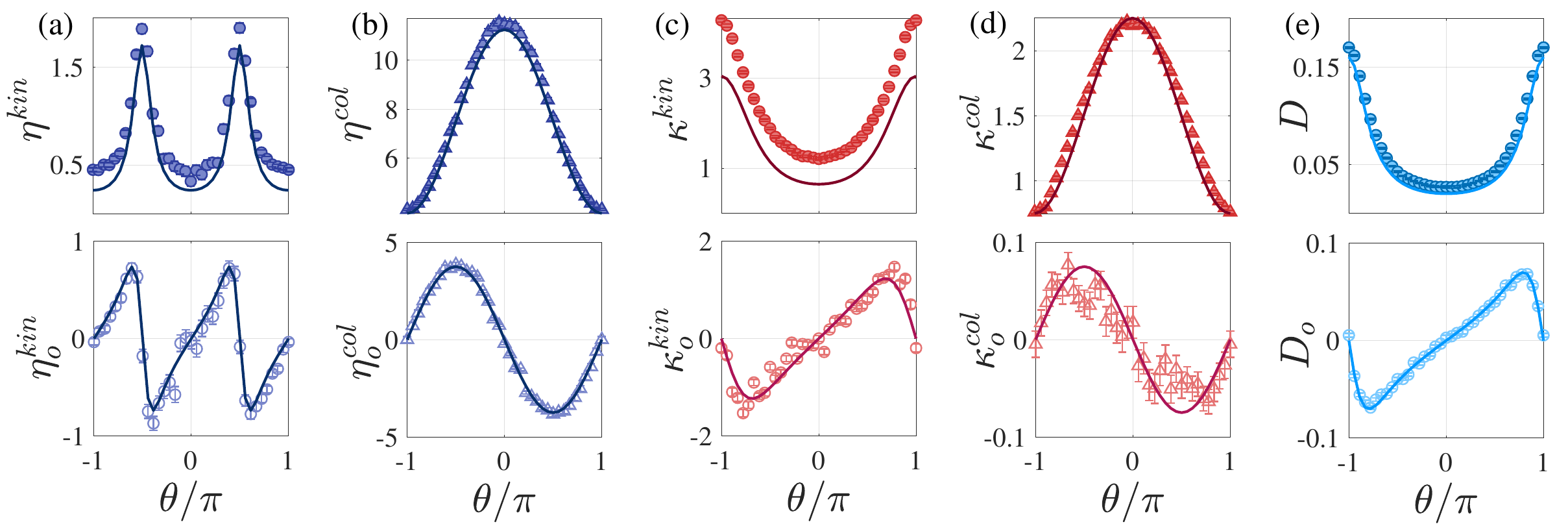}
    \caption{Viscosities (a,b), thermal conductivities (c,d) and self-diffusion coefficients (e) of the CSRD fluid as a function of the additional rotation angle. The solid lines denote the theoretical results calculated from Eqs.~\eqref{Coefficients} and Eqs.~\eqref{SelfDiffusion}, and the symbols are the simulation measurements.  Parameters: $\n=10$ in (a--d), $\n=20$ in (e), and $\omega=2\pi/3$, $\h=0.1$, $k_BT=1$ in (a--e).}
    \label{Fig::Coeff}
\end{figure*}

\section{Hydrodynamic theory and transport coefficients}
The hydrodynamic equations of the CSRD fluid can be derived via a microscopic kinetic theory~\cite{MPC_Kinetic} originally developed by Pooley and Yeomans for determining the transport coefficients of the conventional SRD. In this approach, the system is approximated to be in local equilibrium, with the single-particle distribution being a function of local variables. From the CSRD dynamics (Eqs.~\eqref{strEq}-\eqref{colEq}), the fluxes of conserved quantities, including mass, momentum and energy, are directly derived through lengthy calculations (see Appendices~\ref{APPENDIX::B} and \ref{APPENDIX::C}). Note that both streaming and collision steps may contribute to fluxes of the conserved quantities. The transport coefficients can be extracted from the expressions of these fluxes. And, substituting the fluxes into the corresponding conservation equations separately yields the continuity equation, Navier-Stokes equation and heat conduction equation,
\begin{equation}\label{GenHyEq}
    \begin{aligned}
        &\pd{t}\rho+\pd{\a} \left( \rho u_\a\right)=0,\\
        &\pd{t}\left( \rho u_\a \right)+\pd{\b}\left( \rho u_\a u_\b \right)=\pd{\b}\sigma_{\a\b},\\
       &\frac{k_B}{m}\left[ \pd{t}\left( \rho T \right)+\pd{\a}\left( \rho T u_\a \right)\right]=\sigma_{\a\b}\pd{\b}u_\a+\pd{\a}\left( \kappa_{\a\b}\pd{\b}T \right),
    \end{aligned}
\end{equation}
with $\rho=mn\left( \bm{r},t \right)$ the local mass density ($n$ the particle number density), $\bm{u}\left( \bm{r},t \right)$ the local fluid velocity, and $T\left( \bm{r},t \right)$ the local temperature. Here, the stress tensor $\sigma_{\a\b}$ and the thermal conductivity tensor $\kappa_{\a\b}$ take the forms:
\begin{equation}\label{StrAndHF}
    \begin{aligned}
        \sigma_{\alpha\beta}=&-p\kr{\a\b}+\left[\eta\left(\delta_{\alpha\mu}\delta_{\beta\nu}+\delta_{\alpha\nu}\delta_{\beta\mu}-\delta_{\alpha\beta}\delta_{\mu\nu}\right)\right.\\
        &+\zeta\delta_{\alpha\beta}\delta_{\mu\nu}+\eta_R\left(\delta_{\alpha\mu}\delta_{\beta\nu}-\delta_{\alpha\nu}\delta_{\beta\mu}\right)\\
        &+\eta_o\left(\varepsilon_{\alpha\mu}\delta_{\beta\nu}+\varepsilon_{\beta\nu}\delta_{\alpha\mu}\right)\\
        &-\left.\eta_A\ep{\a\b}\kr{\mu\nu}-\eta_B\kr{\a\b}\ep{\mu\nu}\right]\partial_\nu u_\mu,\\
        \kappa_{\a\b}=&\kappa\kr{\a\b}+\kappa_o\ep{\a\b},
    \end{aligned}
\end{equation}
with $p$ the pressure.

In the stress tensor of Eq.~\eqref{StrAndHF}, $\eta$ refers to the shear viscosity, $\zeta$ to the bulk viscosity, $\eta_R$ to the rotation-rotation viscosity, $\eta_o$ to the odd viscosity, $\eta_A$ to the compression-rotation viscosity, and $\eta_B$ to the rotation-compression viscosity. Their physical meanings can be clearly found for instance in reference~\cite{Vitelli_Fluhydro}. Therein, $\eta$, $\zeta$ and $\eta_R$ are even with respect to $\theta$, while $\eta_o$, $\eta_A$ and $\eta_B$ are odd. Additionally, the thermal conductivity tensor also contains even ($\kappa$) and odd ($\kappa_o$) components. With the contributions separately from the streaming/kinetic ($kin$) and collision ($col$) steps, these coefficients become,
\begin{equation}\label{VisAndCT}
    \begin{aligned}
        &\eta=\eta^{kin}+\frac{1}{2}\eta^{col},\qquad \zeta=\frac{1}{2}\eta^{col},\qquad \eta_R=\frac{1}{2}\eta^{col}\\
        &\eta_o=\eta^{kin}_o+\frac{1}{2}\eta^{col}_o,\qquad \eta_A=-\frac{1}{2}\eta^{col}_o,\qquad \eta_B=\frac{1}{2}\eta^{col}_o\\
        &\kappa=\kappa^{kin}+\kappa^{col},\qquad \kappa_o=\kappa^{kin}_o+\kappa^{col}_o.
    \end{aligned}
\end{equation}
Here, the kinetic and collisional contributions are analytically calculated as,
\begin{widetext}
    \begin{subequations}
        \label{Coefficients}
        \begin{eqnarray}
            &\eta^{kin}&=n\kbT\h\left[\frac{\n}{\n-1+e^{-\n}}\frac{1-\cos2\theta\cos2\omega}{1+\cos2\omega(\cos2\omega-2\cos2\theta)}-\frac{1}{2}\right],\\
            &\eta^{col}&=\frac{m}{12\h}\left( \n-1+e^{-\n} \right)\left( 1-\cos\theta\cos\omega \right),\\
            &\eta_o^{kin}&=-n\kbT\h\frac{\n}{\n-1+e^{-\n}}\frac{\sin2\theta\cos2\omega}{1+\cos2\omega(\cos2\omega-2\cos2\theta)},\\
            &\eta_o^{col}&=\frac{m}{12\h}\left( \n-1+e^{-\n} \right)\sin\theta\cos\omega,\\
            &\kappa^{kin}&=n\frac{k_B^2 T}{m}\h\left\{ \frac{\parbox{11cm}{\centering$\n^2\left[ \left( \n+2\cos\omega \right)\sin^2\frac{\omega}{2}+\left( \left( \n-2 \right)\cos\omega+4\cos2\omega\cos^2\frac{\theta}{2} \right)\sin^2\frac{\theta}{2} \right]$}}{\parbox{11cm}{\centering$\left( \n-1 \right)\left\{ \left( \n+2\cos\omega \right)^2\sin^4\frac{\omega}{2}\right.$\\$\left.+\left[ \left( \n-2 \right)\left( \n-1-\cos2\omega \right)\cos\omega+4\left( \n-1 \right)\cos2\omega\cos^2\frac{\theta}{2} \right]\sin^2\frac{\theta}{2} \right\}$}} -1 \right\}\label{kkin},\\
            &\kappa^{col}&=\frac{k_B\left( \n-1 \right)}{6\n\h}\left( 1-\cos\theta\cos\omega \right),\\
            &\kappa_o^{kin}&=-n\frac{k_B^2 T}{m}\h\frac{\parbox{11cm}{\centering$\n^2\left[ \left( \n-2 \right)\cos\omega+2\cos2\omega\cos\theta \right]\sin\theta$}}{\parbox{11cm}{\centering$2\left( \n-1 \right)\left\{ \left( \n+2\cos\omega \right)^2\sin^4\frac{\omega}{2}\right.$\\$\left.+\left[ \left( \n-2 \right)\left( \n-1-\cos2\omega \right)\cos\omega+4\left( \n-1 \right)\cos2\omega\cos^2\frac{\theta}{2} \right]\sin^2\frac{\theta}{2} \right\}$}},\\
            &\kappa_o^{col}&=\frac{k_B\left( \n-1 \right)}{6\n^2\h}\sin\theta\cos\omega\label{kocol},
        \end{eqnarray}
    \end{subequations}
\end{widetext}
with $\n=nl^2$ the mean number of particles per cell. The molecular chaos assumption has been used for the calculation of viscosities and thermal conductivities. To analytically obtain the thermal conductivities in Eqs.~\eqref{kkin}-\eqref{kocol}, a large density limit ($\n\gg1$) is employed. We point out that the collisional odd thermal conductivity in \eqr{kocol} does not result from a theoretical derivation but an empirical construction, since the molecular chaos hypothesis predicts a vanishing $\kappa_o^{col}$, as detailed in Appendix~\ref{APPENDIX::B}.

From Eqs.~\eqref{GenHyEq}-\eqref{VisAndCT}, the conservation equations of mass, momentum and energy can be written in a concise form,
\begin{equation}\label{simHyEq}
    \begin{aligned}
        &\frac{\mathrm{d}\rho}{\mathrm{d}t}+\rho\nabla\cdot\bm{u}=0,\\
        &\rho\frac{\mathrm{d}\bm{u}}{\mathrm{d}t}=-\nabla p+\hat{\eta}\nabla^2\bm{u}+\hat{\eta}_o \bm{\varepsilon}\cdot\nabla^2\bm{u},\\
        &nk_B\frac{\mathrm{d} T}{\mathrm{d}t}=\bm{\sigma}:\nabla\bm{u}+\kappa\nabla^2 T,
    \end{aligned}
\end{equation}
where $\hat{\eta}=\eta^{kin}+\eta^{col}$ and $\hat{\eta}_o=\eta^{kin}_o+\eta^{col}_o$.
We emphasize that the additional unidirectional rotation (i.e., $\theta$) does not produce a net torque on the fluid. Consequently, the anti-symmetric hydrostatic stress is lacking in the CSRD fluid, different from active spinner fluids that are powered by nonzero torque and have nonvanishing anti-symmetric stress even in a quiescent situation~\cite{Vitelli_Fluhydro,OddEffects1,Lubensky2005}.

Besides the viscosities and thermal conductivities, the self-diffusion coefficients of the CSRD particles are also derived through the same route. Herein, some particles are tagged, such that the CSRD fluid is regarded as a binary mixture (say A and B) with their respective density fields $\rho_A\left( \bm{r},t \right)$ and $\rho_B\left( \bm{r},t \right)$. The density fields evolve according to the isothermal self-diffusion equation,
\begin{equation}\label{EqOfSelfDiffusion}
    \pd{t}\Delta\rho=\pd{\a}D_{\a\b}\pd{\b}\Delta\rho,
\end{equation}
with $\Delta\rho=\rho_A-\rho_B$ and $D_{\a\b}=D\kr{\a\b}+D_o\ep{\a\b}$ the self-diffusion tensor. Based on the diffusive particle flux obtained through the kinetic theory, the normal and odd self-diffusion coefficients ($D$ and $D_o$) separately read,
\begin{equation}\label{SelfDiffusion}
    \begin{aligned}
        D=&\,\frac{\kbT\h}{2m}\frac{2\n\left( 1-\cos\omega\cos\theta \right)}{\left( \n-1+e^{-\n} \right)\left( 1-2\cos\omega\cos\theta+\cos^2\omega \right)}\\
        &-\frac{\kbT\h}{2m},\\
        D_o=&\,\frac{-\n\kbT\h\cos\omega\sin\theta}{m\left( \n-1+e^{-\n} \right)\left( 1-2\cos\omega\cos\theta+\cos^2\omega \right)}.
    \end{aligned}
\end{equation}

Now, we perform simulations to confirm the existence of odd transport coefficients in the CSRD fluid and to verify the validity of the analytical expressions in Eqs.~\eqref{Coefficients} and~\eqref{SelfDiffusion}  (see Appendix~\ref{APPENDIX::A} for the simulation details). In simulations, a boundary-driven shear flow, thermal gradient or concentration gradient is applied to the CSRD fluid, and the resulting steady-state stress, heat flux or particle flux is respectively measured. Thus, the corresponding transport coefficients are accurately quantified, as plotted in Fig.~\ref{Fig::Coeff} that shows the presence of considerable odd transport coefficients. Remarkably, the theoretical predictions based on the kinetic theory are in good agreement with the simulation measurements (Fig.~\ref{Fig::Coeff}). 

\section{Odd fluid transportation}
To see the effects of the odd transport coefficients, three typical transport phenomena in odd fluids are investigated by both simulation and hydrodynamic theory (simulation details are provided in Appendix~\ref{APPENDIX::A}).

\begin{figure}[h]
    \centering
    \includegraphics[keepaspectratio, width=\columnwidth]{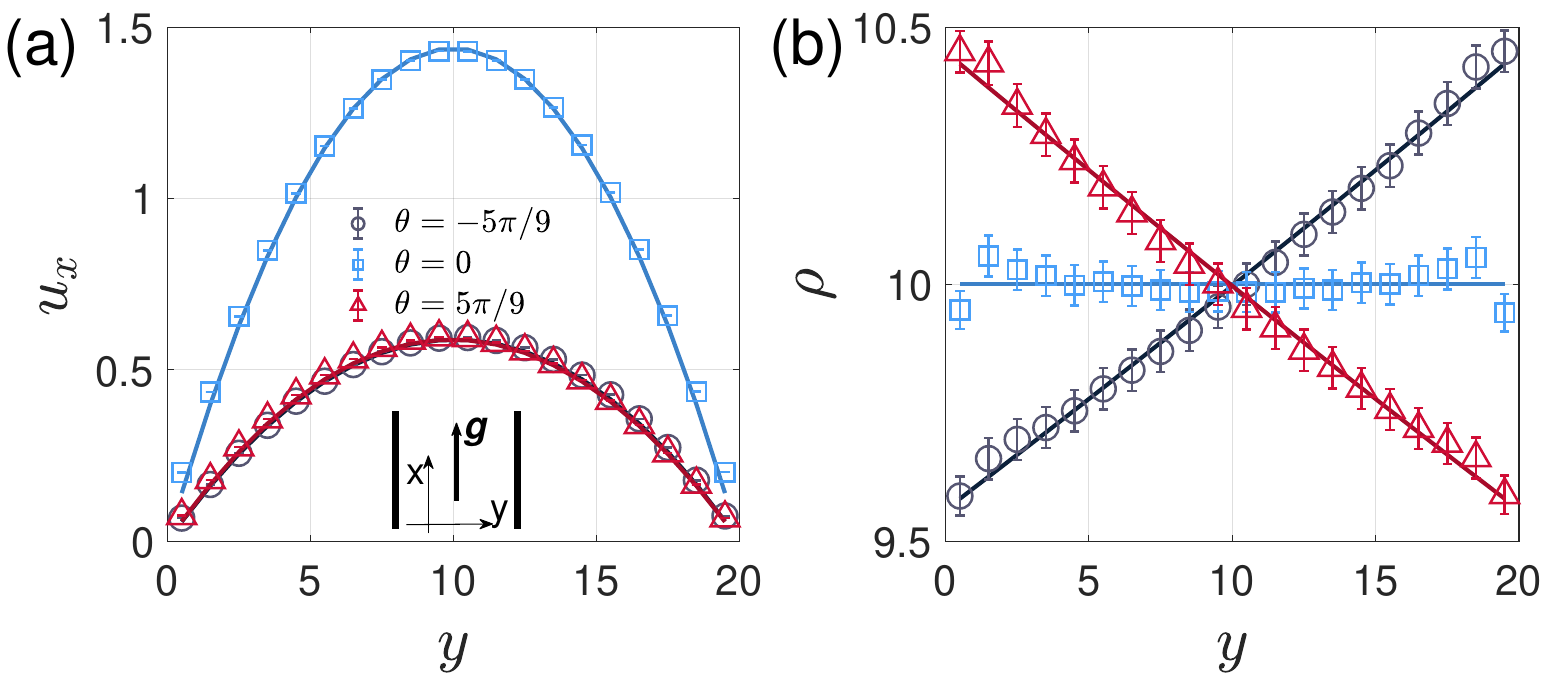}
    \caption{(a) Velocity profile of the fluid flowing through a channel composed of two nonslip walls, and (b) the corresponding density distribution across the channel. The symbols and lines, respectively, denote the results from the CSRD simulations and hydrodynamic theory where the transport coefficients in \eqr{Coefficients} are used. Different symbols represent different $\theta\in(-5\pi/9,0,5\pi/9$). System parameters: $\omega=2\pi/3$, $\h=0.5$, $k_BT=1$, $\rho_0=10$, $g=0.01$, and $L=20$.}
    \label{Fig::Poi}
\end{figure}

The first example considers a fluid flowing through a channel with stick boundary walls, as sketched in Fig.~\ref{Fig::Poi}(a). The fluid is driven by a constant external force $\bm{g}$ and its temperature remains fixed by scaling the particle thermal energy. The simulations reveal that both common fluid ($\theta=0$) and odd fluid exhibit a Poiseuille flow profile (Fig.~\ref{Fig::Poi}(a)). While, unlike the common fluid, the density distribution of the odd fluid is inhomogeneous across the channel and depends on the odd viscosity of the fluid (Fig.~\ref{Fig::Poi}(b)). This finding is consistent with our previous work on active spinner fluid~\cite{OddEffects1}, where the odd viscosity has been proven to cause a transverse Hall-like mass transport under a nonuniform shear. Theoretically, combining Eqs.~\eqref{simHyEq} with boundary condition $\left.\bm{u}(y)\right|_{y=0,L}=\bm{0}$, the solutions of density and velocity field are given by $\rho=\rho_0,\, u_x=\frac{\rho_0 g}{2\hat{\eta}}y\left( L-y \right)$ for $\hat{\eta}_o=0$ and $\rho=\frac{\g \rho_0 L}{\left( e^{\g L} -1 \right)}e^{\g y},\,u_x=\frac{g\rho_0 L}{\g\left( 1-e^{\g L} \right)\hat{\eta}}\left[ e^{\g y}-1-\left( e^{\g L}-1 \right)\frac{y}{L} \right]$ for nonzero $\hat{\eta}_o$. Here, $\rho_0$ denotes the mean mass density, $L$ the channel width, $g=|\bm{g}|$ and $\g=\frac{\hat{\eta}_o mg}{\hat{\eta}\kbT}$. The theoretical calculations perfectly reproduce the simulation results (see Fig.~\ref{Fig::Poi}). 

\begin{figure}[t]
    \centering
    \includegraphics[keepaspectratio, width=\columnwidth]{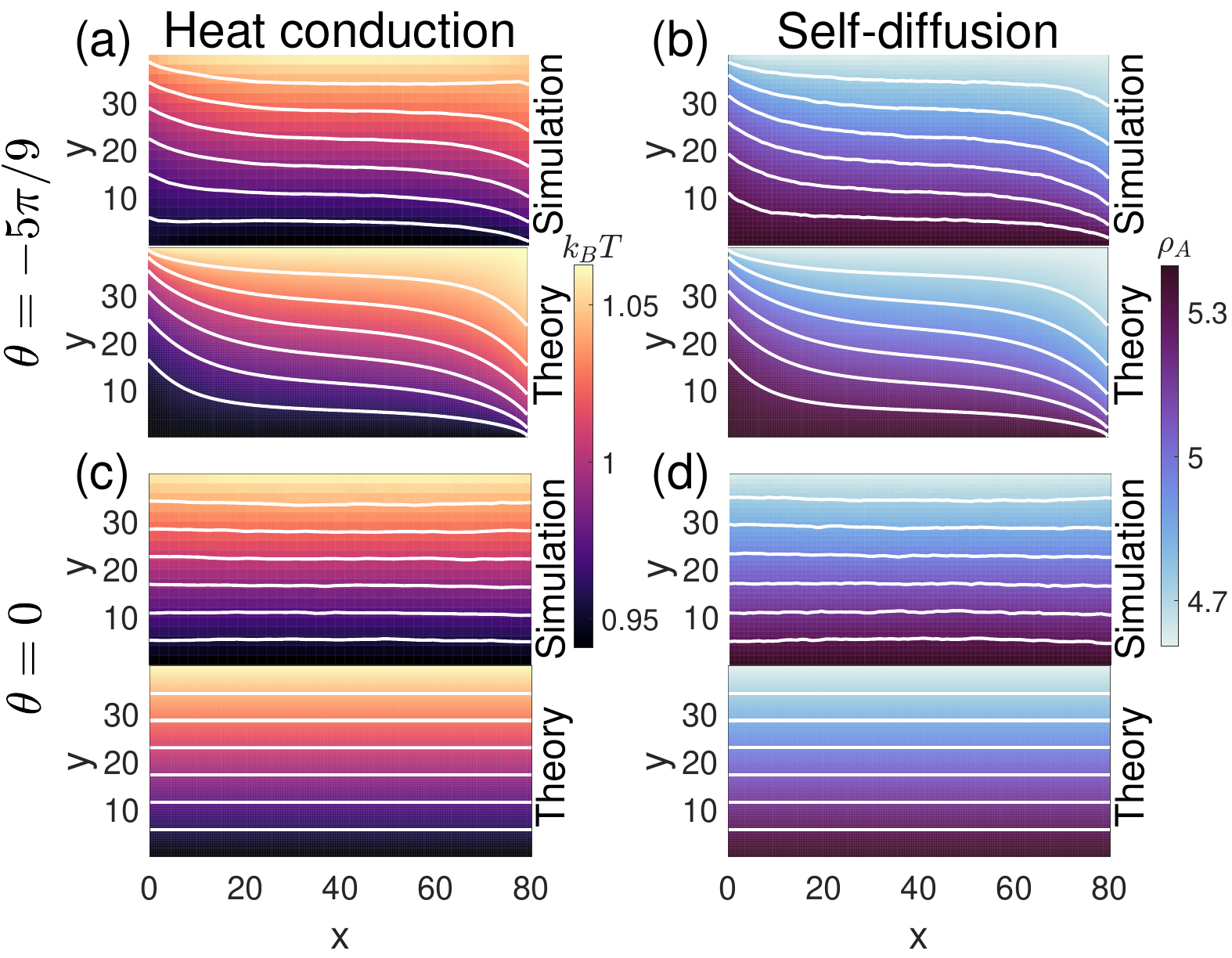}
    \caption{Heat conduction of the odd fluid (a) and common fluid (c) in a confined box, with $\kbT|_{y=0}=0.94$ and $\kbT|_{y=L_y}=1.06$ imposed; self-diffusion of the odd fluid (b) and common fluid (d), with $\rho_A|_{y=0}=5.4$ and $\rho_A|_{y=L_y}=4.6$. Here, the color maps represent the temperature or density distribution, and the white lines are the isotherms or isodensity lines. For comparison, both simulation and theoretical results are given. System parameters: $\h=0.1$, $\n=10$, $\omega=5\pi/6$ for (a,c), $\omega=2\pi/3$ for (b,d), and $\theta=-5\pi/9$ for the odd fluid.}
    \label{Fig::HCSD} 
\end{figure}

The other two examples are steady-state heat conduction and self-diffusion of the odd fluid in a rectangular box of dimensions $L_x\times L_y$. Zero-flux boundary conditions are applied to the left and right walls of the box, while the temperatures (or the tagged-particle density $\rho_A$) at the bottom and top boundaries are fixed to different values for the case of heat conduction (or self-diffusion). The simulations show that the temperature and density distributions of the odd fluid noticeably differ from the common fluid, as presented in Fig.~\ref{Fig::HCSD}. Particularly, in contrast to the common fluids, the isotherms/isodensity lines of the odd fluids are not parallel to the isothermal/isodensity boundary (namely, the temperature or density is not uniform in the $x$ direction). This behavior originates from the transverse heat or particle flux, respectively, induced by the odd thermal conductivity or self-diffusion coefficient.
On the other hand, the continuum theory based on Eqs.~\eqref{simHyEq}-\eqref{EqOfSelfDiffusion} indicates that the steady-state temperature and density fields both obey the Laplace equation, $\nabla^2\phi=0$, with  $\phi=T$ for the heat conduction and $\phi=\rho_A$ for the self-diffusion. With the zero-flux boundaries $\left( \kappa\pd{x}T + \kappa_o\pd{y}T\right)|_{x=0,L_x}=0$ or $\left( D\pd{x}\rho_A + D_o\pd{y}\rho_A\right)|_{x=0,L_x}=0$, and the condition that the values of $\phi|_{y=0, L_y}$ keep fixed, the Laplace equation is solved numerically. The theoretical results compare well to the simulations (Fig.~\ref{Fig::HCSD}), with a relative deviation below $2\%$ (see Appendix~\ref{APPENDIX::A})
, highlighting the validity of the CSRD in describing the transport behaviors of odd fluids.

\section{Odd colloidal suspension}
Since the CSRD inherits all advantages of the traditional SRD, it is immediately applicable to simulating suspensions of diverse mesoscopic objects (e.g., colloids) by utilizing the techniques well developed for the SRD. To demonstrate the capability of the CSRD approach and the exotic behavior of odd complex fluids, we simulate a passive colloidal particle immersed in the odd fluid by applying a hybrid simulation method.

\subsection{Hybrid simulation method for colloidal particles in odd fluids}
For the traditional SRD model, the suspensions of mesoscopic objects are simulated by using the hybrid SRD-MD (molecular dynamics) method~\cite{MPC_MD}. This scheme can be straightforwardly extended to our CSRD model. Therein, the immersed objects and their interactions with the fluid are described by the MD, while the fluid particles are simulated by the CSRD. Without loss of generality, we employ the Weeks-Chandler-Andersen (WCA) potential, $\varphi\left( r \right)=4\epsilon\left[ \left( \frac{R}{r} \right)^{12}-\left( \frac{R}{r} \right)^{6} +\frac{1}{4} \right]$ for $\quad r\leqslant 2^{1/6}R$, to model the interaction between the CSRD particles and colloidal particle, with $\epsilon$ the interaction intensity and $R$ the colloidal radius.

During the streaming step, the colloidal and CSRD particles evolve by solving the Newtonian equations of motion via the velocity-Verlet algorithm with a time step $dt=\Delta t/50$:
\begin{equation*}
    \begin{aligned}
        &\bm{r}_i\left( t+dt\right) = \bm{r}_i\left( t \right) + \bm{v}_i\left( t \right)dt + \frac{\bm{F}_i\left( t \right)}{2m}dt^2,\\
        &\bm{r}^p\left( t+dt \right) = \bm{r}^p\left( t \right) + \bm{v}^p\left( t \right)dt + \frac{\bm{F}^p}{2M^p}dt^2,\\
        &\bm{v}_i\left( t+dt\right) = \bm{v}_i\left( t \right) + \frac{dt}{2m}\left( \bm{F}_i\left( t \right) + \bm{F}_i\left( t+dt\right)\right),\\
        &\bm{v}^p\left( t+dt\right) = \bm{v}^p\left( t \right) + \frac{dt}{2M^p}\left( \bm{F}^p\left( t \right) + \bm{F}^p\left( t+dt\right)\right),
    \end{aligned}
\end{equation*}
where $\bm{F}_i$ and $\bm{F}^p$ refer to the force on fluid particle $i$ and that on the colloidal particle due to the WCA potential, and $M^p$, $\bm{r}^p$, and $\bm{v}^p$ are the mass, position and velocity of the colloidal particle, respectively. Actually, for most fluid particles that are far away from the colloidal particle (i.e. $\bm{F}_i=0$), their streaming steps just reduce to Eq.~(\ref{strEq}). The CSRD collision step remains unchanged, where the fluid particles update their velocities according to Eq.~(\ref{colEq}).

Because the WCA potential only provides radial forces normal to the surface of colloidal particle, the tangential components of the fluid velocity do not change during the collision with the colloidal particle. As a result, the WCA potential leads to a slip colloidal particle. In order to realize the no-slip boundary, a coarse-grained bounce-back rule is additionally applied to the fluid particles that collide with the colloidal particle after the velocity-Verlet MD update. 
During the bounce-back operation, the velocity of the fluid particle relative to the colloidal particle surface is reversed, which ensures that the average relative velocity of the fluid particles near the colloidal surface is zero. The post-collision velocities of the fluid particles and colloidal particle can be determined by using the conservation of momentum, angular momentum and energy during the bounce-back collision. For simplicity, here we do not consider the rotation of the colloidal particle (which is equivalent to a colloidal particle with the mass $M^p\gg m$), and the post-collision velocities reduce to,
\begin{equation}
   \begin{aligned}
        \bm{v}_i'&=-\left( \bm{v}_i - \bm{v}^p \right) + {\bm{v}^p}',\\
        {\bm{v}^p}'&=\bm{v}^p + 2\frac{m}{mN_b+M^p}\sum_i\left( \bm{v}_i - \bm{v}^p \right).
    \end{aligned}
\end{equation}
Here the superscript prime denotes the velocities after the bounce-back collision, $N_b$ is the number of fluid particles participating in the interaction with the colloid, and the summation runs for all such fluid particles. We point out that these coarse-grained no-slip and slip boundary conditions are not only highly effective for our current mesoscale simulations but meet our simulation requirements appropriately. While, developing a precise microscopic theory for the hydrodynamic boundary conditions of the mesoscale odd fluid remains a valuable pursuit.

Our following colloidal simulations are performed in an $L\times L$ box with periodic boundary conditions. Additionally, the Maxwell-Boltzmann scaling thermostat, a thermostat developed for the traditional SRD~\cite{MBS_Huang}, is used to maintain an isothermal condition. When investigating the flow past the colloidal particle, we fix the colloidal particle at the box center and generate the fluid flow by endowing the CSRD particles inside the bin of $L-1<y<L$ with velocities sampled from the Maxwell distribution with a mean velocity $\bm{u}_\infty$ and the temperature $k_BT$. Simulation parameters used in colloidal simulation are: $\omega=2\pi/3$, $\Delta t=0.1$, $dt=0.002$, $k_BT=1$, $\lambda=10$, $R=4$, $\epsilon=2.5k_BT$, $F_y=2$, $L=80$, $\bm{u}_\infty=0.05\bm{e}_y$, and $M^p=\lambda\pi R^2$. The resulting Reynolds number is about 0.2. 

\subsection{Flow field and colloidal dynamics}
\begin{figure}[t]
    \centering
    \includegraphics[keepaspectratio, width=\columnwidth]{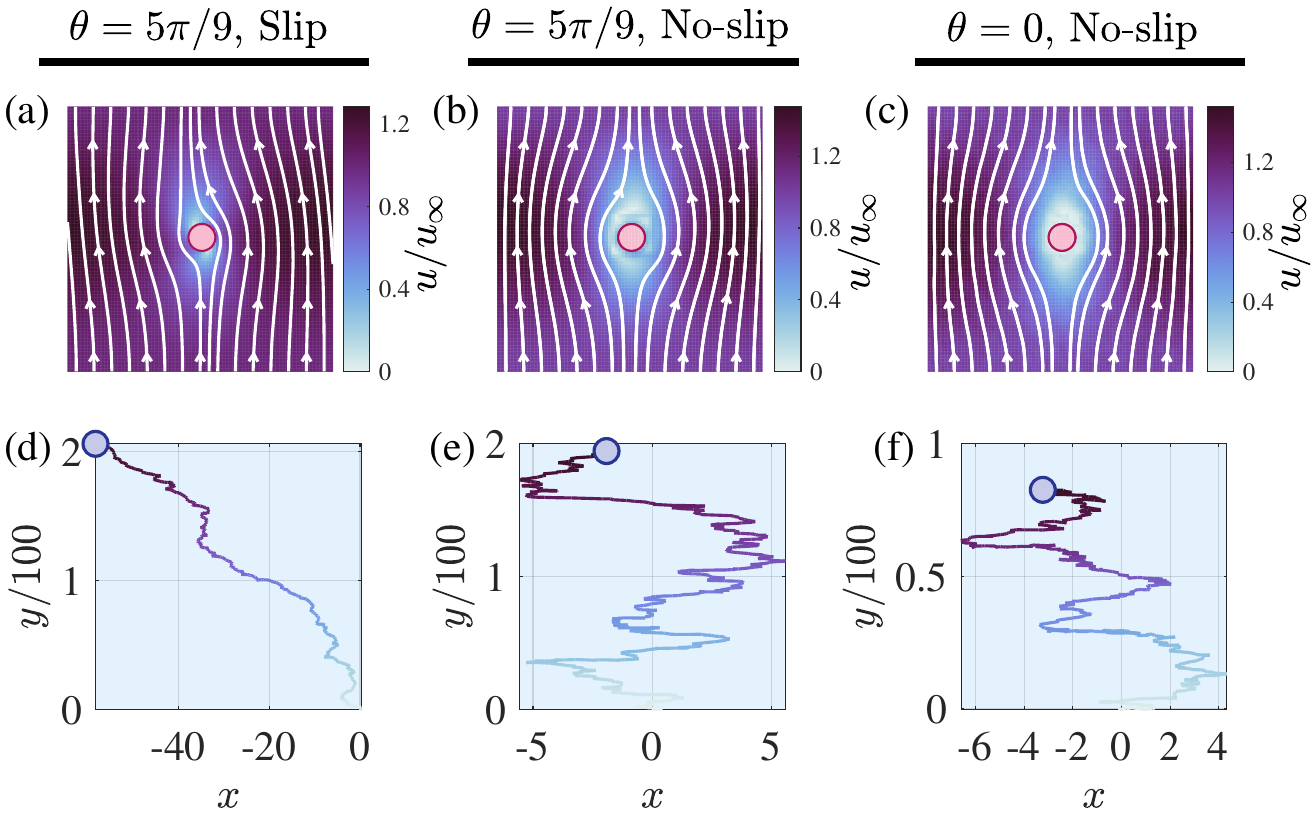}
    \caption{{(a--c) Velocity field of the fluid flowing upward through a fixed colloidal particle with the white streamlines (color indicates the relative flow speed). (d--f) Trajectories of the colloidal particle migrating under a vertical downward external force.}}
    \label{Fig::Col} 
\end{figure}

To begin with, we consider an odd/normal fluid flowing past a fixed colloidal particle with slip or no-slip boundary. The simulation results are depicted in Fig.~\ref{Fig::Col} (for the normal fluid ($\theta=0$), we only provide the no-slip result since it is qualitatively similar to the slip case). Different from the normal fluid (Fig.~\ref{Fig::Col}(c)), the flow pattern of the odd fluid around the slip colloidal particle (Fig.~\ref{Fig::Col}(a)) shows nonsymmetrically distorted streamlines, consistent with the recent theoretical prediction~\cite{Lier2024}; while the odd fluid flow around the no-slip particle (Fig.~\ref{Fig::Col}(b)) is similar to the case of the normal fluid. The reason for this is that in the incompressible limit (valid in our simulations), the odd viscosity term in the Navier-Stokes equation can be absorbed into the pressure, $\rho\frac{\mathrm{d}\bm{u}}{\mathrm{d}t}=-\nabla\tilde{p}+\hat{\eta}\nabla^2\bm{u}$, with $\tilde{p}=p+\hat{\eta}_o\ep{\a\b}\pd{\a}u_\b$. Consequently, the flow pattern can only be influenced by the odd viscosity via the boundary conditions involving the stress, e.g. the slip boundary.

\begin{table}[h]
    \centering
    \caption{Even and odd mobilities, $\mu^p$ and $\mu^p_o$, and even and odd self-diffusion coefficients, $D^{p}$ and $D_o^{p}$ of colloidal particle.\label{Tab::Mobility}}
    \begin{tabular}{c|cccc}
    \hline\hline
                         & \multicolumn{2}{c}{$\theta=0$} & \multicolumn{2}{c}{$\theta=9\pi/5$} \\ \cline{2-5} 
                         &   Slip    & Noslip  &   Slip    & No-slip  \\ \hline
    $\mu^{p}k_BT$       & 0.0129(1)& 0.0095(1) & 0.0241(2) & 0.0201(1)\\
    $D^{p}$         & 0.0129(3)& 0.0092(3) & 0.023(1)  & 0.020(1)\\
    $\mu_o^{p}k_BT$     &    0      &     0     &-0.0058(2) &    0     \\
   $D_o^{p}$        &    0      &     0     &-0.0058(1) &    0     \\ \hline\hline
    \end{tabular}
\end{table}

We further investigate the dynamics of a slip or no-slip colloidal particle migrating in the CSRD/SRD fluid under an external force $\bm{F}=F_y\bm{e}_y$. The trajectories of the colloidal particle are shown in Fig.~\ref{Fig::Col}(d--f). We note that a transverse migration appears in the case of the slip colloidal particle in the odd fluid (Fig.~\ref{Fig::Col}(d)), significantly different from other situations (Figs.~\ref{Fig::Col}(e) and (f)). From the stationary migrating velocity $U_\a$, the mobility tensor ${\mu}_{\a\b}^{p}=\mu^p\kr{\a\b}+\mu_o^p\ep{\a\b}$ of the colloidal particle is quantified by the relation $U_\a=\mu_{\a\b}^pF_\b$. The results are given in Table.~\ref{Tab::Mobility}, showing the slip colloidal particle in the odd fluid has a nonvanishing odd mobility $\mu_o^{p}$. {Finally, we determine both normal and odd components of the colloidal diffusion tensor $D_{\a\b}^p=D^p\kr{\a\b}+D_o^p\ep{\a\b}$ by measuring the velocity correlations of a free colloidal particle~\cite{Hargus2021} via $D_{\a\b}^p=\int_0^\infty\mathrm{d}t\left\langle v_\a^p\left( t \right) v_\b^p\left( 0 \right) \right\rangle$. The diffusion tensor and the corresponding mobility tensor are found to satisfy the Einstein relation $D_{\a\b}^p=\mu_{\a\b}^pk_BT$, aligning with the equilibrium nature of the CSRD.}

\section{Conclusion}
We develop an efficient mesoscale simulation approach for odd fluids and analytically derive its transport coefficients and continuum equations. The mesoscopic odd fluid not only exhibits all odd transport coefficients but also properly captures the unusual dynamics of odd fluids/complex fluids. Our work thus paves the way for simulating odd complex fluids at large scales and exploring their unique characteristics.

\begin{acknowledgments}
This work was supported by the National Key R\&D Program of China (2022YFF0503504), National Natural Science Foundation of China (No. T2325027, No. 12274448).
\end{acknowledgments}
\appendix

\renewcommand\thefigure{A\arabic{figure}}  
\renewcommand\thetable{A\arabic{table}}  
\renewcommand{\theequation}{A\arabic{equation}}
\setcounter{equation}{0}
\setcounter{figure}{0}
\setcounter{table}{0}
\section{Simulation details}
\label{APPENDIX::A}
In this appendix, we provide the simulation details for the measurement of the transport coefficients and for the case studies in the main text.

\subsection{Transport coefficients}
The transport coefficients are computed by performing nonequilibrium simulations. First, a gradient of velocity, temperature, or concentration is applied to the system. To generate these gradients, a method named reverse nonequilibrium molecular dynamics (RNEMD), that was originally proposed by Florian M\"uller-Plathe~\cite{RNEMD_veloctiy,RNEMD} for determining transport coefficients of fluids in molecular dynamics simulation, is utilized. In this method, the gradient and hence its corresponding flux are created by using the Maxwell's demon-type operation to change the attributes of particles on the system boundaries. This boundary operation does not influence physical processes in the bulk. Then, the steady-state gradient and (longitudinal and transverse) fluxes are measured. Finally, according to the constitutive relations between the gradient and flux, the transport coefficients are obtained.

\subsubsection{\textbf{Viscosities}}
As shown in the main text, all of the viscosities are the combinations of $\eta^{kin}, \eta^{col}, \eta_o^{kin},$ and $\eta_o^{col}$. These viscosities can be calculated by imposing a simple shear with the shear rate $\dot{\gamma}=\pd{y}u_x$. According to Eqs.~(4) and (5) in the main text, the stress in the presence of this shear rate is given by:
\begin{equation}\label{StressShear}
    \begin{aligned}
        &\s_{xx}=-p+\eta_o^{kin}\dot{\gamma},\qquad \s_{yy}=-p-\left( \eta_o^{kin}+\eta_o^{col} \right)\dot{\gamma},\\
        &\s_{xy}=\left( \eta^{kin}+\eta^{col} \right)\dot{\gamma},\qquad \s_{yx}=\eta^{kin}\dot{\gamma},
    \end{aligned}
\end{equation}
where $p$ is the pressure. The stress also consists of the kinetic and collision parts: $\s_{\a\b}=\s_{\a\b}^{kin}+\s_{\a\b}^{col}$. Each part of the stress can be read from Eq.~\eqref{StressShear}. Referring to the derivations in Appendix.~\ref{SEC:Momentumflux}, the pressure term only comes from the kinetic contribution, so the kinetic part of the stress is:
\begin{equation}\label{StressShearKin}
    \begin{aligned}
        &\s_{xx}^{kin}=-p+\eta_o^{kin}\dot{\gamma},\qquad \s_{yy}^{kin}=-p-\eta_o^{kin}\dot{\gamma},\\
        &\s_{xy}^{kin}=\eta^{kin}\dot{\gamma},\qquad \s_{yx}^{kin}=\eta^{kin}\dot{\gamma}.
    \end{aligned}
\end{equation}
And, the collision part of the stress is:
\begin{equation}\label{StressShearCol}
    \begin{aligned}
        &\s_{xx}^{col}=0,\qquad \s_{yy}^{col}=-\eta_o^{col}\dot{\gamma},\\
        &\s_{xy}^{col}=\eta^{col}\dot{\gamma},\qquad \s_{yx}^{col}=0.
    \end{aligned}
\end{equation}
Therefore, the viscosities $\eta^{kin}, \eta^{col}, \eta_o^{kin},$ and $\eta_o^{col}$ can be calculated from Eqs.~\eqref{StressShearKin} and \eqref{StressShearCol}:
\begin{equation}
    \begin{aligned}
        &\eta^{kin}=\frac{\s_{xy}^{kin}+\s_{yx}^{kin}}{2\dot{\gamma}},\qquad \eta_o^{kin}=\frac{\s_{xx}^{kin}-\s_{yy}^{kin}}{2\dot{\gamma}},\\
        &\eta^{col}=\frac{\s_{xy}^{col}}{\dot{\gamma}},\qquad \eta_o^{col}=-\frac{\s_{yy}^{col}}{\dot{\gamma}}.
    \end{aligned}
\end{equation}

The simple shear with $\dot{\gamma}=\pd{y}u_x$ is generated by the RNEMD approach (see \cite{RNEMD_veloctiy} for more details). Briefly, the CSRD simulation system, with the size $L$ and the periodic boundary conditions, is divided into identical slabs along the $y$ axis. The width of slabs is $l$, equal to the size of the collision cell. Pick out the particle with the minimum $x$-component momentum, $p_{x}^{(-)}$ (i.e., the maximum value against the $x$ axis) in the bottom slab (the region $\left\{ (x,y)\mid 0\leqslant y\leqslant l \right\}$) and the particle with the largest $x$-component momentum, $p_{x}^{(+)}$, in the middle slab (the region $\left\{ (x,y)\mid L/2 \leqslant y\leqslant L/2+l \right\}$) every $W$ CSRD steps. Then, the momentums $p_x$ of these two particles are exchanged. The exchange interval $W$ is large enough to avoid a non-linear response. In the simulation here, we set $L=20l$ and $W=40$. After a quick relaxation, the system reaches the steady state and a symmetric linear shear field (with respect to $y=(L + l)/2$) is established. It should be pointed out that the total energy of the system remains constant in the RNEMD-type simulation. 

The momentum flux ($T_{\a\b}= \rho u_\a u_\b-\s_{\a\b}$) across an imaginary plane (locating in the lower or upper half simulation box) can be directly measured. Its kinetic part $T_{\a\b}^{kin}=\rho u_\a u_\b-\s_{\a\b}^{kin}$ is quantified by accumulating the momentum $p_\a$ of the particles crossing the plane, with the normal vector along the $\b$ direction, in the streaming step. The collision part of the momentum flux $T_{\a\b}^{col}=-\s_{\a\b}^{col}$ is obtained by measuring the momentum transfer across the plane induced by the rotational collision. 

\subsubsection{\textbf{Thermal conductivities}}
If a temperature gradient $\pd{y}T$ is imposed in the CSRD fluid, the thermal conductivities $\kappa^{kin}, \kappa^{col}, \kappa_o^{kin},$ and $\kappa_o^{col}$ can be determined by the following relations,
\begin{equation}
    \begin{aligned}
        &q_x^{kin}=-\kappa_o^{kin}\pd{y}T,\qquad q_y^{kin}=-\kappa^{kin}\pd{y}T,\\
        &q_x^{col}=-\kappa_o^{col}\pd{y}T,\qquad q_y^{col}=-\kappa^{col}\pd{y}T.
    \end{aligned}
\end{equation}
Here $\bm{q}^{kin}$ and $\bm{q}^{col}$ are the heat fluxes from the kinetic and collision steps, respectively.

In the simulation, the temperature gradient is created as follows. Every $W$ simulation steps, the particle with the smallest kinetic energy and that with the largest kinetic energy are selected out from the bottom and middle slabs, respectively. If the kinetic energy of the former is smaller than the latter, their velocities will be swapped. Such velocity-swap operation results in a continuous heat flux and hence thermal gradient in the system, which is also symmetric with respect to $y=(L + l)/2$. To produce a linear thermal gradient, the exchange interval is taken as $W=40$ in the simulation. After the system reaches the steady state, the heat fluxes contributed due to the streaming and collision steps can be separately measured.

\subsubsection{\textbf{Self-diffusion coefficients}}

The self-diffusion flow obeys the relation $J^D_\a=-D_{\a\b}\pd{\b}\Delta\rho$, with $\Delta\rho=\rho_A-\rho_B$ and $\bm{J}^D=\bm{J}^A-\bm{J}^B$. Note that particles of species A and B are identical, and they are only distinguished by an imagined tag. Under the gradient $\pd{y}\Delta\rho$, the self-diffusion coefficients $D_{\a\b}=D\kr{\a\b}+D_o\ep{\a\b}$ are computed by the following equations:
\begin{equation}\label{I-6}
    J_x^{D}=-D_o\pd{y}\Delta\rho,\qquad J_y^{D}=-D\pd{y}\Delta\rho.
\end{equation}
\par
The gradient $\pd{y}\Delta\rho$ is obtained by performing an artificial ``chemical reaction'' for the particles in the bottom and middle slabs. This chemical reaction transforms the particle $A$ to $B$ with probability $p_{AB}$ and the particle $B$ to $A$ with probability $p_{BA}=1-p_{AB}$. For the bottom slab, the probability $p_{BA}$ is set to be larger than $p_{AB}$, so that more $A$ particles are produced there. While in the middle slab, $p_{BA}$ is smaller than $p_{AB}$. This reaction takes place also every $W=40$ CSRD steps. In the initial state, $A$ and $B$ particles are uniformly distributed with the same density $\rho_A=\rho_B$, i.e. $\Delta\rho=0$. In the simulation, we take $p_{AB}=0.46$ in the bottom slab and $p_{AB}=0.54$ in the middle slab. In the steady state, the concentration gradient and self-diffusion flux $\bm{J}^D$ are directly measured, such that the self-diffusion coefficients are extracted from Eq.~\eqref{I-6}. 

\subsection{Case studies}

\subsubsection{\textbf{Poiseuille flow in a channel}}

Under a gravity field $\bm{g}=\left( g,0 \right)$, the CSRD fluid develops a Poiseuille flow along the $x$ direction in a channel with length $L_x=20l$ and width $L_y=20l$. The channel has the periodic boundary condition in the $x$ direction and no-slip boundary walls in the $y$ direction realized by the bounce-back collision. The gravity force acts on the particles in the streaming step through the following equations,
\begin{equation}
\begin{aligned}
 r_{\a,i}(t+\h)&=r_{\a,i}(t)+v_{\a,i}(t)\h+\frac{1}{2}g_\a\h^2,\\
 v_{\a,i}(t+\h)&=v_{\a,i}(t)+g_\a\h. 
\end{aligned}
\end{equation}
The system is coupled to a thermostat to achieve an isothermal flow. Specifically, every $W_t=2$ simulation steps, the particles are sorted into the cells of a fixed square lattice with size $2l$. To avoid confusion, these cells are called as the thermostat cell (TC). The instant local temperature $k_BT_{c}$ of a TC containing $N_{c}$ particles ($N_{c}>1$) reads,
\begin{equation}
 k_BT_c=\frac{1}{2(N_c-1)}\sum_{i\in TC} \left( \bm{v}_i - \bm{v}_{c} \right)^2,
\end{equation}
where $\bm{v}_c=\frac{1}{N_c}\sum_{i\in TC}\bm{v}_i$ is the center-of-mass velocity of this TC. In order to maintain the temperature at a desired value of $k_BT_0$, the velocities of the particles, relative to the center of mass, are rescaled:
\begin{equation}
    \bm{v}_i^\pr=\bm{v}_c+\left( \bm{v}_i-\bm{v}_c \right)\sqrt{\frac{k_BT_0}{k_BT_c}},\qquad i\in TC
\end{equation}
where $\bm{v}_i^\pr$ is the velocity after thermostatting.

Theoretically, from the hydrodynamic equations of the CSRD fluid (Eqs.~(7) in the main text) and the ideal-gas equation of state, the Poiseuille flow velocity field driven by $\bm{g}$ obeys the following equations,
\begin{equation}\label{PoiEq}
    \begin{aligned}
        \hat{\eta}\pd{y}^2u_x&=-\rho g,\\
        \hat{\eta}_o\pd{y}^2u_x&=-\frac{k_BT}{m}\pd{y}\rho,
    \end{aligned}
\end{equation}
with the boundary conditions $\left.u_x(y)\right|_{y=0,L}=0$ and $\int_0^Ldx\int_{0}^{L}dy\rho/L^2=\rho_0$. Thus, the flow field and density distribution can be solved analytically.

\subsubsection{\textbf{Heat conduction in a confined box}}

In the simulation, the CSRD fluid is confined in a $L\times L$ box with $L=80l$. The boundaries in the $x$ direction are reflective walls, while the periodic boundary condition is applied in the $y$ direction. The RNEMD method is used to produce a thermal gradient. Here, the system is divided into slabs with width $l$ along the $y$ axis, and the kinetic energy of the particles in the bottom and middle slabs are swapped every $W=40$ steps. Because of the periodic boundary condition in the $y$ direction, the resulted temperature distribution is symmetric with respect to the position $y=L/2$, and the bottom (and top) and middle slabs separately have the minimum and maximum steady-state temperatures, which are $k_BT(y=0)=0.94$ and $k_BT(y=L/2)=1.06$, respectively. Due to the system symmetry, we only present the simulation results for the lower half simulation box. 

In theory, this heat conduction problem is described by the Laplace equation,
\begin{equation}
    \nabla^2 T=0,
\end{equation}
with the boundary conditions,
\begin{equation}
    \begin{aligned}
        &\left.q_x\right|_{x=0}=\left.-\left( \kappa\pd{x} T + \kappa_o\pd{y} T\right)\right|_{x=0}=0,\\
        &\left.q_x\right|_{x=80l}=\left.-\left( \kappa\pd{x} T + \kappa_o\pd{y} T\right)\right|_{x=80l}=0,\\
        &T(y=0)=0.94,\qquad T(y=40l)=1.06.
    \end{aligned}
\end{equation}
The above equation is numerically solved by finite difference method (FDM), where the thermal conductivities $\kappa,\,\kappa_o$ are given by Eqs.~(6e)-(6h) in the main text. Figures~\ref{Fig:SDHCSM}.(a,c) display the results from the CSRD simulation and theoretical calculation, and the relative error between them (defined as $|Theory-Simulation|/Simulation$).

\subsubsection{\textbf{Self-diffusion in a confined box}}

The simulation setup for the self-diffusion is similar to the case of heat conduction in last section. Here, the particles in the bottom slab, no matter which type of particles they are, are re-labeled as $A$ with probability $p_A=0.54$ and as $B$ with probability $p_B=1-p_A=0.46$. While, in the middle slab, the particles are re-labeled as $A$ with probability $p_A=0.46$ and as $B$ with probability $p_B=0.54$. This re-labeling operation results in fixed concentrations for both species on the boundaries. And the system is also symmetric about $y=L/2$. In the steady state, the densities of species $A$ in the bottom and middle slabs are, respectively, $\rho_A(y=0)=5.4$ and $\rho_A(y=L/2)=4.6$.

\begin{figure}[h]
    \centering
    \includegraphics[keepaspectratio, width=1\columnwidth]{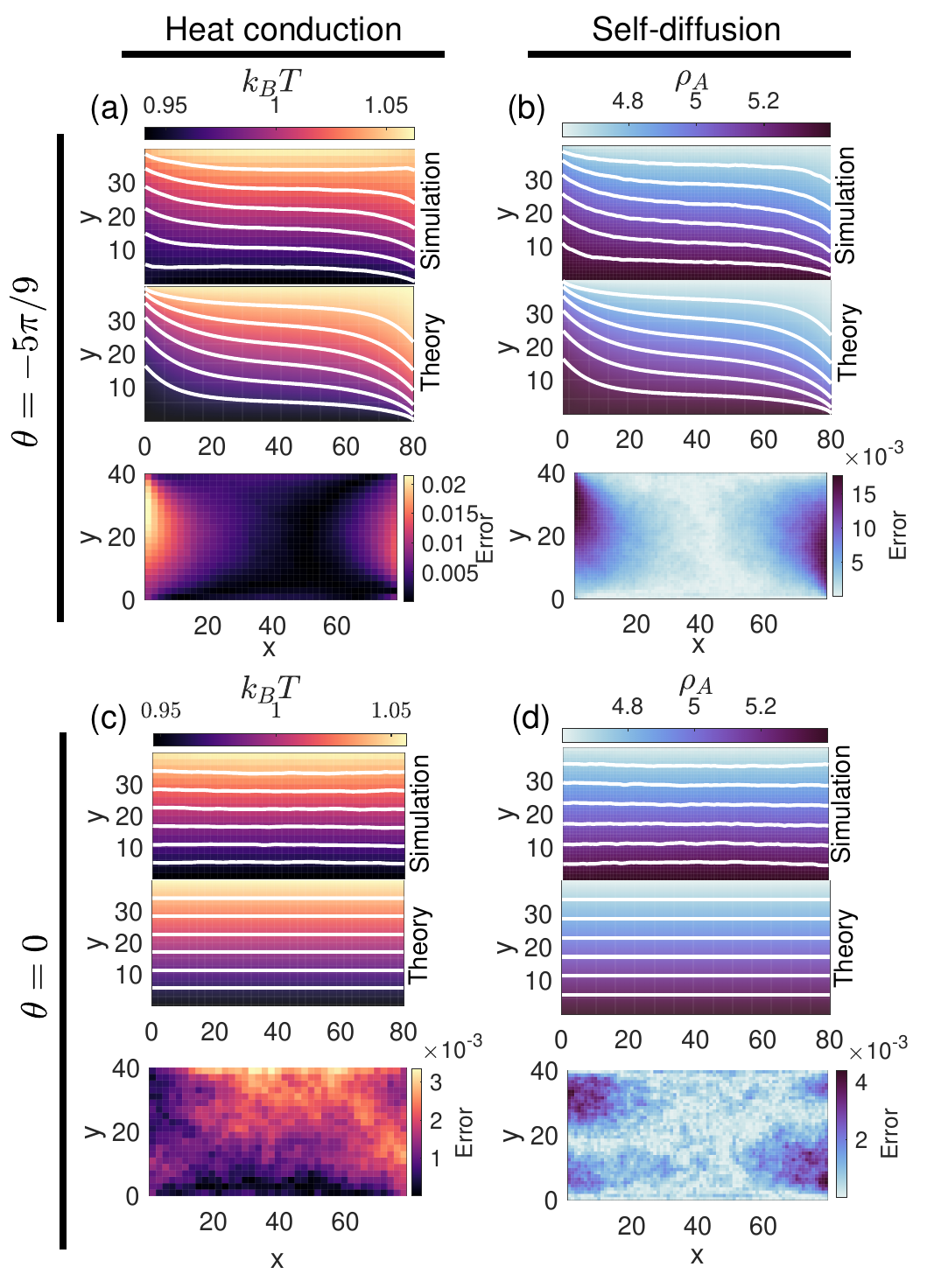}
    \caption{Temperature maps in heat conduction (a,c) and density maps in self-diffusion  (b,d) of the CSRD fluid in a confined box. System parameters: $\h=0.1$ and $\n=10$ for (a-d), $\omega=5\pi/6$ for (a,c), $\omega=2\pi/3$ and $k_BT=1$ for (b,d), $\theta=-5\pi/9$ for (a,b) and $\theta=0$ for (c,d). Here, the white lines are the isotherms or isodensity lines. The lowest panels in (a-d) refer to the relative error between the simulation and theoretical results.}
    \label{Fig:SDHCSM}
\end{figure}

Theoretically, the equation and boundary conditions satisfied by the diffusion problem read,
\begin{equation}
    \nabla^2 \rho_A=0.
\end{equation}
\begin{equation}
    \begin{aligned}
        &\left.J^D_x\right|_{x=0}=-2\left.\left( D\pd{x}\rho_A + D_o\pd{y}\rho_A\right)\right|_{x=0}=0,\\
        &\left.J^D_x\right|_{x=80l}=-2\left.\left( D\pd{x}\rho_A + D_o\pd{y}\rho_A\right)\right|_{x=80l}=0,\\
        &\rho_A(y=0)=5.4,\qquad \rho_A(y=40l)=4.6.
    \end{aligned}
\end{equation}
This boundary value problem of the Laplace equation is solved by the FDM, in which the coefficients $D$ and $D_o$ are given by Eqs.~(9) in the main text. For comparison, the simulation and theoretical results are shown in Figs.~\ref{Fig:SDHCSM}(b,d).

\renewcommand\thefigure{B\arabic{figure}}  
\renewcommand\thetable{B\arabic{table}}  
\renewcommand{\theequation}{B\arabic{equation}}
\setcounter{equation}{0}
\setcounter{figure}{0} 
\setcounter{table}{0}
\section{Theoretical derivation of transport coefficients and hydrodynamic equations}\label{APPENDIX::B}
\subsection{Assumptions and definitions}

A non-equilibrium kinetic theory originally proposed by Pooley and Yeomans in \cite{MPC_Kinetic} is applied to derive the hydrodynamic equations and transport coefficients of the CSRD fluid. The basic idea and definitions employed in this theory are introduced below.

To begin with, the particle number density is assumed to be large enough so that it is appropriate to treat the CSRD fluid as a continuous medium. The single-particle distribution function is denoted as $f=f(\bm{r},\bm{v})$ with normalization condition $\int d\bm{r}d\bm{v}f=m\mathcal{N}$, where $m$ is the mass of a particle and $\mathcal{N}$ is the total particle number. Then the mass density and momentum distribution is obtained by $\rho(\bm{r})\triangleq\IntV f$ and $\rho(\bm{r})\bm{u}(\bm{r})\triangleq\IntV\bm{v}f$ respectively, with $\bm{u}(\bm{r})$ the flow velocity field. The distribution of an arbitrary quantity $X=X(\bm{r},\bm{v})$ is defined by $\bk{X(\bm{r})}\triangleq\frac{1}{\rho}\IntV Xf$~\footnote{In this article, $\bk{\cdot}$ is used to represent the average on velocity and $\ave{\cdot}$ for the average on any other stochastic variables.}. The moment of velocity is defined as $M_{\a\b\ldots}\triangleq\bk{\left( v_\a-u_\a \right)\left( v_\b-u_\b \right)\cdots}$. The temperature is defined according to the second moment of velocity, $\T\triangleq k_BT/m=\frac{1}{2}M_{\a\a}$ (setting $k_B=1$)~\footnote{Unless otherwise stated, the repeated indices are represented the summation and the Greek alphabet subscript are denoted for the components of coordinates $x,\,y$.}.

It is assumed that the CSRD system under study is in local equilibrium, so that $f(\bm{r},\bm{v})$ is the local equilibrium distribution and is a function of the local thermodynamic quantities($\rho(\bm{r}),\T(\bm{r}),\ldots$):
\begin{equation}\label{localEqdistribution}
f(\bm{r},\bm{v})=\frac{\rho(\bm{r})}{\T(\bm{r})}g\left( \frac{\bm{v}-\bm{u}(\bm{r})}{\sqrt{\T(\bm{r})}} \right),
\end{equation}
where $g(x)$ is a function of the dimensionless quantity $x$. Under the hydrodynamic limit, the thermodynamic quantities fluctuate at the large time and space scales, such that their gradients can be treated as small quantities, whose magnitude is denoted as $\delta$.

The hydrodynamic equations of the CSRD fluid can be derived from the conservation equation of the conserved quantity $Q$ that is the mass, momentum, and energy. The general form of the conservation equation of $Q$ is,
\begin{equation}\label{Q}
\pd{t}\rho_Q+\pd{\a}J_\a^{(Q)}=0,
\end{equation}
where $\rho_Q$ and $\bm{J}^{(Q)}$ are the density and the flux of $Q$, respectively. Following the method used in \cite{MPC_Kinetic}, the continuity equation, the Navier-Stokes equation and the heat conduction equation can be obtained by calculating the specific form of flux $\bm{J}^{(Q)}$, when $Q$ takes $m$, $m\bm{v}$ and $\frac{1}{2}mv^2$ separately. 

Because the evolution of the CSRD is discrete in time, the flux, which is related to the quantity crossing a plane during the time interval $\left[ t,t+\h \right]$, corresponds to the ``discrete flux'' $\bm{j}^{(Q)}$  rather than the ``continuous flux'' $\bm{J}^{(Q)}$ associated with continuous time dynamics. The expression of ``discrete flux'' $\bm{j}^{(Q)}$ is,
\begin{equation}
j_\a^{(Q)}(t)=\frac{1}{\h}\int_{t}^{t+\h}dt^\prime J_\a^{(Q)}(t^\prime).
\end{equation}
According to the mean value theorem of integrals, the above formula is expressed by $j_\a^{(Q)}(t)=J_\a^{(Q)}(t+\tau),\tau\in\left[ 0,\h \right]$, which allows representing  $J_\a^{(Q)}$ by $j_\a^{(Q)}$: $J_\a^{(Q)}(t)=j_\a^{(Q)}(t-\tau)$. Then, expanding $j_\a^{(Q)}(t-\tau)$ at $t$ to the order of $\mathcal{O}(\delta)$ yields,
\begin{equation}
J_\a^{(Q)}(t)=j_\a^{(Q)}(t)-\tau\pd{t}j_\a^{(Q)}(t)+\mathcal{O}(\delta^2).
\end{equation}
Substitute this into \eqr{Q} and set $\tau=\frac{1}{2}\h$ for approximation, the conservation equation of the CSRD fluid is obtained as follows,
\begin{equation}\label{SRDQ}
\pd{t}\rho_Q+\pd{\a}\left( j_\a^{(Q)}-\frac{\h}{2}\pd{t}j_\a^{(Q)} \right)=\mathcal{O}(\delta^3).
\end{equation}
The flux, $j_\a^{(Q)}$, originates from both streaming and collision steps, which can be derived separately. In the streaming step, $j_\a^{(Q)}$ is proportional to the mean quantity of $Q=Q(\bm{v})$ flowing across an imaginary plane (line in 2D) of unit area during $\h$, which is denoted as $\Delta Q$. In general, $\Delta Q$ depends on the distribution function $f(\bm{r},\bm{v})$. With the help of the expansion of $f(\bm{r},\bm{v})$ at a point on the plane, $\Delta Q$ can be approximated by the velocity moments at that point. Using the updated rule of the CSRD, an iterative equation for the velocity moment is derived. The fixed point of this equation gives the value of the velocity moment, and from this value, the form of the flux in the streaming step is obtained. While, the flux contributed from the collision step is directly derived by calculating the average transfer of $Q$ through the plane that separates the collision cell into two compartments, because the cell-level conservation of $Q$ is obeyed during the collision. Subsequently, substituting $j_\a^{(Q)}$ into the conservation equation \eqr{SRDQ} yields the hydrodynamic equations and transport coefficients. In the remaining sections, the derivation of $j_\a^{(Q)}$ is provided in detail.

\,\,
\subsection{The calculation of the discrete flux: the general form}
\subsubsection{\textbf{The flux in the streaming step}}

The flux $j_\a^{(Q)}$ in the streaming step, denoted by $j_\a^{(Q),kin}$ can be written down directly in terms of its definition. For example, consider a line segment $\ell=\left\{ \left( x,y \right) |\left|x\right|\leq\frac{a}{2},y=0 \right\}$ without loss of generality so that the flux $j_y^{(Q),kin}$ across it during $\h$ is,
\begin{equation}\label{jy}
\begin{aligned}
    j_y^{(Q),kin}&=\frac{1}{a\h}\IntV\int_{-v_y\h}^{0}dr_y\int_{-\frac{a}{2}+\frac{v_x}{v_y}r_y}^{\frac{a}{2}+\frac{v_x}{v_y}r_y}dr_xQ(\bm{v})f/m\\
    &\triangleq\frac{1}{a\h}\IntV\int_{A}d\bm{r}Q(\bm{v})f/m.
\end{aligned}
\end{equation}
This integration can be calculated by expanding the distribution function $f$ with respect to the origin point to the order of $\mathcal{O}(\delta)$, in which the expansions of the density, flow velocity and temperature, are utilized,
\begin{align}
\rho&=\rho_0+r_\a \bko{\pd{\a}\rho}+\mathcal{O}(\delta^2),\\
u_\b&=u_{0\b}+r_\a \bko{\pd{\a}u_\b}+\mathcal{O}(\delta^2),\\
\T&=\T_0+r_\a \bko{\pd{\a}\T}+\mathcal{O}(\delta^2),
\end{align}
with the subscript $0$ denoting the value taken at the origin point. Here, $u_{0\b}$ is considered to be $\mathcal{O}(\delta)$ because the Galilean invariance of the CSRD allows to choose such a reference frame. Thus, using the following substitution
\begin{equation}\label{VChange}
\frac{\bm{v}^\pr-\bm{u}_0}{\sqrt{\T_0}}=\frac{\bm{v}-\bm{u}}{\sqrt{\T}}
\end{equation}
in the definition \eqr{localEqdistribution} and ignoring $\mathcal{O}(\delta^2)$ order quantities, \eqr{jy} becomes
\begin{equation}\label{Kinjy}
\begin{aligned}
    j_y^{(Q),kin}=\frac{1}{a\h}\IntV^\pr  &\int_{A^\pr}d\bm{r}f(\bm{0},\bm{v}^\pr)\\
    &\cdot\frac{1}{m}Q\left(\bm{v}\right)\left[ 1+\frac{1}{\rho_0}r_\a\bko{\pd{\a}\rho} \right],
\end{aligned}
\end{equation}
where $\bm{v}=\left( 1+\frac{1}{2\T_0}r_\a\bko{\pd{\a}\T} \right)\bm{v}^\pr+r_\a\bko{\pd{\a}\bm{u}}$, $A^\pr=\left\{ (r_x,r_y)|r_x^l\leq r_x\leq r_x^u,r_y^l\leq r_y\leq 0 \right\}$ and the limits of integration $r_x^l,r_x^u,r_y^l$ are
\begin{widetext}
    \begin{equation}
        \begin{aligned}
            r_x^l&=-\frac{a}{2}+\frac{v_x^\pr}{v_y^\pr}\left[ 1+\left( \frac{\pdo{x}{u_x}}{v_x^\pr}-\frac{\pdo{x}{u_y}}{v_y^\pr} \right)\left( -\frac{a}{2}+\frac{v_x^\pr}{v_y^\pr}r_x \right)+\left( \frac{\pdo{y}{u_x}}{v_x^\pr}-\frac{\pdo{y}{u_y}}{v_y^\pr} \right)r_y \right]r_y,\\
            r_x^u&=\frac{a}{2}+\frac{v_x^\pr}{v_y^\pr}\left[ 1+\left( \frac{\pdo{x}{u_x}}{v_x^\pr}-\frac{\pdo{x}{u_y}}{v_y^\pr} \right)\left( \frac{a}{2}+\frac{v_x^\pr}{v_y^\pr}r_x \right)+\left( \frac{\pdo{y}{u_x}}{v_x^\pr}-\frac{\pdo{y}{u_y}}{v_y^\pr} \right)r_y \right]r_y,\\
            r_y^l&=-v_y^\pr\h\left[ 1+\left( \frac{\bko{\pd{x}\T}}{2\T_0}+\frac{\bko{\pd{x}u_y}}{2v_y^\pr} \right)r_x-\left( \frac{\bko{\pd{y}\T}}{2\T_0}-\frac{\bko{\pd{y}u_y}}{2v_y^\pr} \right)v_y^\pr\h \right].
        \end{aligned}
    \end{equation}        
\end{widetext}

The integral over $\bm{r}$ in \eqr{Kinjy} can be implemented by expressing $Q(\bm{v})$ in terms of $\bm{v}^\pr$. To the order of $\mathcal{O}(\delta)$, only the integrations like
$\int_{A^\pr} d\bm{r}$ and $\int_{A^\pr} d\bm{r}r_\a$ need to be calculated and their results are given by \cite{MPC_Kinetic}:
\begin{equation}\label{Area0}
\begin{aligned}
        A\triangleq&\, \frac{1}{a\h}\int_{A^\pr} d\bm{r}\\
        =&\,v_y^\pr-\h\left\{v_y^\pr\pdo{y}{u_y}+\frac{1}{2}\left[ v_y^\pr\pdo{x}{u_x} + v_x^\pr\pdo{x}{u_y} \right]\right.\\
        &\left.+\frac{1}{2\T_0}\left[ \pdo{y}{\T}{v_y^\pr}^2+\pdo{x}{\T}v_x^\pr v_y^\pr \right]\right\},
\end{aligned}
\end{equation}
\begin{equation}\label{Area1}
R_\a\triangleq\int_{A^\pr} d\bm{r}r_\a=-\frac{1}{2}a\h^2v_y^\pr v_\a^\pr.
\end{equation}
As a consequence, the integration \eqr{Kinjy} has the form of $\IntV^\pr f(\bm{0},\bm{v}^\pr)\mathcal{P}(\bm{v}^\pr)$ after integrating $\bm{r}$ using \eqr{Area0} and \eqr{Area1}, in which $\mathcal{P}(\bm{v}^\pr)$ is a polynomial of $\bm{v}^\pr$. Hence, \eqr{Kinjy} can generally be represented by the first, second, and third-order moments of the velocity at the origin point. 

\subsubsection{\textbf{The flux in the collision step}}
It is convenient to calculate the ``collision flux'' in a single collision cell. We use the superscript $col$ to label it, i.e., $j_\a^{(Q),col}$. The center of this cell is set as the origin point and the domain of the cell is defined as $C=\left\{ (r_x,r_y)|\left|r_x\right|\leq\frac{l}{2},\left|r_y\right|\leq\frac{l}{2} \right\}$ where $l=1$ is the length of the cell. Imaging a fixed horizontal line segment $r_y=c$ that crosses the collision cell, the flux $j_y^{(Q),col}$ across this line segment contributed by the collision can be written as
\begin{equation}\label{Coljy}
\begin{aligned}
        &j_y^{(Q),col}\\
        &=\ave{\frac{1}{l\h}\int_{-l/2}^{l/2}dr_x\int_{c}^{l/2}dr_y\rho\bk{Q\left( \bm{v}^c \right)/m-Q\left( \bm{v} \right)/m}},
\end{aligned}
\end{equation}
with $\bm{v}$ and $\bm{v}^c$ being the velocities of a particle before and after the collision, respectively. Here, the lower limit of the integral, $r_y=c$, is a random variable uniformly distributed between $-l/2$ and $l/2$, which originates from the random shift operation before performing the collision. It is worth noting that the collision step transfers the momentum and kinetic energy instead of the mass.

\subsection{The calculation of the mass flux}
Substituting $Q=m$ into \eqr{Kinjy} and combining with \eqr{Area0} and \eqr{Area1}, the mass flux $j_\a^{(m)}$ to the order of $\mathcal{O}(\delta)$ is
\begin{equation}\label{jm}
\begin{aligned}
        &j_\a^{(m)}=j_\a^{(m),kin}\\
        &=\rho_0u_{0\a}-\frac{1}{2}\h\pdo{\a}{\rho\T}-\frac{1}{2}\h\pdo{\b}{\rho u_\b u_\a}+\mathcal{O}(\delta^2).
\end{aligned}
\end{equation}
The subscript $0$ in \eqr{jm} can be ignored to represent the mass flux at arbitrary point. Then according to the conservation equation \eqr{SRDQ}, the continuity equation is obtained as follows,
\begin{equation}
    \begin{aligned}
        \pd{t}\rho+\pd{\a} \biggl\{ \rho u_\a &-\frac{\h}{2}\left[ \pd{t}\left( \rho u_\a \right) \right.\\
        &\left.+ \pd{\b}\left( \rho u_\b u_\a \right) + \pd{\a}\left( \rho\T \right) \right] \biggr\}=\mathcal{O}(\delta^3).
    \end{aligned}
\end{equation}
This equation can be simplified further by using a conclusion obtained in the derivation of momentum flux (see \eqr{NSeq}): $\pd{t}\left( \rho u_\a \right)+ \pd{\b}\left( \rho u_\b u_\a \right) + \pd{\a}\left( \rho\T \right)=\mathcal{O}(\delta^2)$, from which the standard form of continuity equation is obtained,
\begin{equation}
\pd{t}\rho+\pd{\a} \left( \rho u_\a\right)=\mathcal{O}(\delta^3)=0.
\end{equation}

\subsection{The calculation of the momentum flux}\label{SEC:Momentumflux}
Define the momentum flux density tensor $T_{\a\b}\triangleq j_\b^{(p_\a)}, T_{\a\b}=T_{\a\b}^{kin}+T_{\a\b}^{col}$, where $T_{\a\b}^{kin}=j_\b^{(p_\a),kin}$ and $T_{\a\b}^{col}=j_\b^{(p_\a),col}$. The fluxes $T_{\a\b}^{kin}$ and $T_{\a\b}^{col}$ are derived as follows.

\subsubsection{\textbf{The momentum flux in the streaming step}}
The general form of $T_{\a\b}^{kin}$ at arbitrary point can be derived by inserting $Q=m\bm{v}$ into \eqr{Kinjy} and using \eqr{Area0} and \eqr{Area1},
\begin{equation}\label{KinTab}
    \begin{aligned}
        T_{\a\b}^{kin}=\rho&\biggl[ u_\a u_\b+M_{\a\b}\\
        &-\frac{1}{2}\T\h\left( \pd{\a}u_\b+\pd{\b}u_\a+\kr{\a\b}\pd{\gamma}u_\gamma \right) \biggr]+\mathcal{O}(\delta^2).
    \end{aligned}
\end{equation}
Here, the second-order moment of velocity $M_{\a\b}$ will be calculated afterwards.

\subsubsection{\textbf{The momentum flux in the collision step}}
To derive the momentum flux due to the collision step, we select a representative particle whose pre-collision velocity is $\bm{v}$ and let $\hat{\bm{v}}$ be the mean velocity of other particles. Thus, the center of mass velocity of this cell becomes $\bm{v}_{cm}=\frac{1}{N}\bm{v}+\frac{N-1}{N}\hat{\bm{v}}$, with $N\geqslant1$ the particle number in the cell. Here, the particle number in a collision cell approximately obeys the Poissonian distribution, with the mean value of $\n=nl^2$ ($n$ the particle number density). From this, given that there already exists a representative particle inside a cell, the probability that the cell contains $N=q$ particles is $P(N=q)=e^{-\n}\n^{q-1}/(q-1)!,q\geqslant1$.

According to the collision operation in the cell $\bm{\xi}$, the velocity of the representative particle after the collision, $\bm{v}^c$, reads
\begin{equation}\label{colEq_SM}
\bm{v}^c=\bm{v}_{cm}(t)+\bm{R_\xi}\left( \bm{v}(t)-\bm{v}_{cm}(t) \right),
\end{equation}
where $\bm{R_\xi}=\cos\a\bm{I}-\sin\a\bm{\varepsilon}$, with $\bm{I}$ the unit tensor, $\bm{\varepsilon}$ the Levi-Civita tensor and $\alpha=\Omega+\theta$ the rotation angle defined in the Eq.(2) of the main text. Substituting the decomposition $\bm{v}_{cm}=\frac{1}{N}\bm{v}+\frac{N-1}{N}\hat{\bm{v}}$ into the above equation gives,
\begin{equation}\label{colDetailEq}
\begin{aligned}
    v_\a^c=\frac{N-1}{N}\biggl[ &\left( \cos\a+\frac{1}{N-1} \right)v_\a-\sin\a v_\a^\ast\\
    &+\left( 1-\cos\a \right)\hat{v}_\a+\sin\a\hat{v}_\a^\ast\biggr],
\end{aligned}
\end{equation}
where $v_\a^\ast=\ep{\a\b}v_\b$ and $\hat{v}_\a^\ast=\ep{\a\b}\hat{v}_\b$.

The collision-induced change in the local flow velocity can be obtained by taking the average of \eqr{colDetailEq} with respect to $\bm{v}$, $\hat{\bm{v}}$, $N$, the random rotation angle $\Omega$, and the particle positions except for the selected particle, 
\begin{equation}\label{DeltaU}
\begin{aligned}
    \ave{\bk{v_\a^c-v_\a}}&=g_c\left( u_\a-u_{0\a} \right)+g_s\left( u_\a^\ast-u_{0\a}^\ast \right)\\
                  &=g_c\bko{\pd{\gamma}u_\a}r_\gamma+g_s\bko{\pd{\gamma}u_\a^\ast}r_\gamma,
\end{aligned}
\end{equation}
where,
\begin{equation}
\begin{aligned}
    g_c&=1-\frac{1}{\n}\left( \n-1+e^{-\n} \right)\left( 1-\cos\theta\cos\omega \right),\\
    g_s&=-\frac{1}{\n}\left( \n-1+e^{-\n} \right)\sin\theta\cos\omega.
\end{aligned}
\end{equation}
In this derivation, the expansion of local flow velocity about the cell center, $u_\b=u_{0\b}+r_\a \bko{\pd{\a}u_\b}+\mathcal{O}(\delta^2)$, has been employed.

Substituting \eqr{DeltaU} into \eqr{Coljy}, together with $Q\left( \bm{v}^c \right)/m-Q\left( \bm{v} \right)/m=v_\a^c-v_\a$, gives the momentum flux in the collision step at an arbitrary point (removing the subscript $0$),
\begin{equation}\label{ColTab}
T_{\a\b}^{col}=-\eta^{col}\pd{\b}u_\a-\eta_o^{col}\pd{\b}u_\a^\ast.
\end{equation}
Here, $\eta^{col}$ and $\eta_o^{col}$ are, respectively, the shear viscosity and odd viscosity contributed by the collision,
\begin{equation}
\eta^{col}=\frac{m}{12\h}\left( \n-1+e^{-\n} \right)\left( 1-\cos\theta\cos\omega \right),
\end{equation}
\begin{equation}
\eta_o^{col}=\frac{m}{12\h}\left( \n-1+e^{-\n} \right)\sin\theta\cos\omega.
\end{equation}

\subsubsection{\textbf{The Navier-Stokes equation with undetermined moments}}
Putting the total flux $T_{\a\b}=T_{\a\b}^{kin}+T_{\a\b}^{col}$ into \eqr{SRDQ}, we have the momentum conservation equation,
\begin{equation}\label{NS0}
\pd{t}\left( \rho u_\a \right)+\pd{\b}\left( T_{\a\b}-\frac{1}{2}\h\pd{t}T_{\a\b}^{kin} \right)=\mathcal{O}(\delta^3).
\end{equation}
The time derivative of $T_{\a\b}^{kin}$ is
\begin{equation}
\begin{aligned}
    \pd{t}T_{\a\b}^{kin}&=\kr{\a\mu}\kr{\b\nu}\pd{t}\left(\rho M_{\mu\nu}  \right)+\mathcal{O}(\delta^2)\\
    &=\kr{\a\mu}\kr{\b\nu}\pd{t}\left[\rho \left( \T\kr{\a\b}+M_{\a\b}^\pr \right)  \right]+\mathcal{O}(\delta^2)\\
    &=\kr{\a\b}\pd{t}\left( \rho\T \right)+\mathcal{O}(\delta^2),
\end{aligned}
\end{equation}
where a decomposition $M_{\a\b}=\T\kr{\a\b}+M_{\a\b}^\pr$ has been used, with $M_{\a\b}^\pr$ the traceless part of $M_{\a\b}$. The traceless moment $M_{\a\b}^\pr$ can be written in a more symmetric way,
\begin{equation}\label{MabPri}
M_{\a\b}^\pr=M_{1}\bm{\s}_{\a\b}^1+M_{3}\bm{\s}_{\a\b}^3
\end{equation}
where $M_{1}=\frac{1}{2}M_{\a\b}\bm{\sigma}_{\a\b}^1=M_{xy}$, $M_{3}=\frac{1}{2}M_{\a\b}\bm{\sigma}_{\a\b}^3=\frac{1}{2}\left( M_{xx}-M_{yy} \right)$, and the matrices $\bm{\s}_{\a\b}^i$ ($i=0,1,2,3$) are defined as
\begin{equation}
\bm{\s}^0=\bm{I},\quad
\bm{\s}^1=\begin{bmatrix}
    0 & 1 \\
    1 & 0
\end{bmatrix},\quad
\bm{\s}^2=-\bm{\varepsilon},\quad
\bm{\s}^3=\begin{bmatrix}
    1 & 0 \\
    0 & -1
\end{bmatrix}.
\end{equation}

The Navier-Stokes equation can be achieved by plugging the fluxes \eqr{KinTab} and \eqr{ColTab} into \eqr{NS0} and using the energy conservation equation (to the order of $\mathcal{O}(\delta)$) obtained in the next section (see \eqr{ThCoEq}, i.e., $\pd{t}\left( \rho\T \right)=-2\rho\T\pd{\a}u_\a+\mathcal{O}(\delta^2)$),
\begin{equation}\label{NSeq}
\pd{t}\left( \rho u_\a \right)+\pd{\b}\left( \rho u_\a u_\b \right)=\pd{\b}\sigma_{\a\b}.
\end{equation}
with the stress tensor,
\begin{equation}\label{Stress0}
\begin{aligned}
    \sigma_{\a\b}=&\,\sigma_{\a\b}^{kin}+\sigma_{\a\b}^{col},\\
    \sigma_{\a\b}^{kin}=&\,-p\kr{\a\b}+\frac{1}{2}\rho\T\h\left( \pd{\a}u_\b+\pd{\b}u_\a-\kr{\a\b}\pd{\gamma}u_\gamma \right)\\
     &-\rho M_{\a\b}^\pr,\\
    \sigma_{\a\b}^{col}=&\,\eta^{col}\pd{\b}u_\a+\eta_o^{col}\pd{\b}u_\a^\ast.
\end{aligned}
\end{equation}
Here, $p=\rho\T$ is the pressure and the traceless $M'_{\a\b}$ will be calculated in Section~\ref{SEC:2ndMoment}. 

\subsection{The calculation of the energy flux}
Define the energy flux $q_\a\triangleq j_\a^{(E_k)}, q_\a=q_\a^{kin}+q_\a^{col}$, where $E_k=\frac{1}{2}mv_\b v_\b$, $q_\a^{kin}=j_\a^{(E_k),kin}$ and $q_\a^{col}=j_\a^{(E_k),col}$. Their expressions are provided in the remaining of this section.

\subsubsection{\textbf{The energy flux in the streaming step}}
The general expression of $q_\a^{kin}$ can be calculated from \eqr{Kinjy}, \eqr{Area0} and \eqr{Area1}:
\begin{equation}\label{Kinqa}
q_\a^{kin}=2\rho\T u_\a+\frac{1}{2}\rho M_{\b\b\a}-\h\pd{\a}\left(\rho\Tp{2}\right),
\end{equation}
where the third-order velocity moment $M_{\b\b\a}$ will be determined in Section~\ref{SEC:F}.

\subsubsection{\textbf{The energy flux in the collision step}}
The variation of the local energy in \eqr{Coljy} $\ave{\rho\bk{Q\left( \bm{v}^c \right)/m-Q\left( \bm{v} \right)/m}}=\rho\ave{\bk{\left( \bm{v}^c \right)^2-\bm{v}^2}}$, induced by the collision step, can be calculated from \eqr{colDetailEq}. Here, $\left( \bm{v}^c \right)^2-\bm{v}^2$ is derived as
\begin{equation}\label{DeltaV2}
\begin{aligned}
    \left( \bm{v}^c \right)^2-\bm{v}^2=&\,-2\frac{N-1}{N^2}\left( 1-\cos\a \right)\left[ \bm{v}^2-\left( N-1 \right)\hat{\bm{v}}^2 \right]\\
                                      &-2\frac{\left( N-1 \right)\left( N-2 \right)}{N^2}\left( 1-\cos\a \right)\bm{v}\cdot\hat{\bm{v}}\\
                                      &+2\frac{N-1}{N}\sin\a\bm{v}\cdot\hat{\bm{v}}^\ast.
\end{aligned}
\end{equation}
Then, averaging over the particle positions except for the selected particle and the velocities of all the particles under the molecular chaos assumption, the variation of $\bm{v}^2$ is
\begin{equation}\label{AveV2}
\begin{aligned}
    &\ave{\bk{\left( \bm{v}^c \right)^2-\bm{v}^2}}=\ave{-4\frac{N-1}{N^2}\left( 1-\cos\a \right)}\left( \T-\T_0 \right)\\
    &=\ave{-4\frac{N-1}{N^2}\left( 1-\cos\a \right)}r_\a\bko{\pd{\a}\T}+\mathcal{O}(\delta^2).
\end{aligned}
\end{equation}
Substituting this into \eqr{Coljy} and omitting the subscript $0$ yields the energy flux contributed due to the collision,
\begin{equation}\label{colheat}
\begin{aligned}
    q_\a^{col}=&\,\ave{\frac{1}{l\h}\int_{-l/2}^{l/2}dr_x\int_{c}^{l/2}dr_y\frac{1}{2}\rho\bk{\left[ \left( \bm{v}^c \right)^2-\bm{v}^2 \right]}}\\
               =&\,-\frac{m\n}{6\h}\ave{\frac{N-1}{N^2}}\left( 1-\cos\theta\cos\omega \right)\pd{\a}\T\\
               \triangleq&\,-\frac{m}{k_B}\kappa^{col}\pd{\a}\T=-\kappa^{col}\pd{\a}T,
\end{aligned}
\end{equation}
with $\kappa^{col}$ the collision thermal conductivity. Generally, the term $\ave{\frac{N-1}{N^2}}$ cannot be calculated exactly. However, in the large $\n$ limit, $N$ can be simply replaced by $\n$, such that $\kappa^{col}$ becomes,
\begin{equation}
\kappa^{col}=\frac{k_B\left( \n-1 \right)}{6\n\h}\left( 1-\cos\theta\cos\omega \right).
\end{equation}
{Notice that the collision part of the odd thermal conductivity $\kappa_o^{col}$ (which relates heat flux by $q_\a^{col}=-\kappa^{col}\pd{\a} T-\kappa_o^{col}\pd{\a}^\ast T$) does not appear here. This is because under molecular chaos hypothesis used in Eq.~\eqref{AveV2}, only the first term of \eqref{DeltaV2} appears in the energy flux Eq.~\eqref{colheat}. This term is even in $\theta$ ($\mathbb{E}\left[ \cos\alpha \right]=\cos\omega\cos\theta$), so that only the even collisional thermal conductivity appears under this hypothesis. }However, we find the following empirical equation can well describe the odd thermal conductivity for large $\n$,
\begin{equation}
\kappa_o^{col}=\frac{k_B\left( \n-1 \right)}{6\n^2\h}\sin\theta\cos\omega.
\end{equation}
{To construct this empirical form of the collisional odd thermal conductivity $\kappa_o^{{col}}$ in the large-particle-number limit, we proceed as follows.}
\begin{enumerate}
    \item First, $\kappa_o^{{col}}$ is an odd function of $\theta$ and its period is $2\pi$ according to the simulation result, implying $\kappa_o^{{col}} \propto \mathbb{E}\left[ \sin\a \right]$.

    \item Second, the energy flux in a cell during the collision step must vanish when only a single particle exists in the cell (see Eq.~\eqref{DeltaV2}). This requires $\kappa_o^{{col}} \propto \mathbb{E}\left[ (N-1)\sin\a \right]$.

    \item Third, since the collisional thermal conductivity increases with collision frequency, we write:\\
     $\kappa_o^{{col}} \propto \frac{1}{\h} \mathbb{E}\left[ (N-1)\sin\a \right]=\frac{1}{\h}(\lambda-1)\cos\omega\sin\theta$.

    \item Fourth, $\kappa_o^{col}$ arises from the two particle velocity correlation which is a smaller quantity compared to $\kappa^{col}$ (and it is true according to the simulation results). From this, we assume $\frac{\kappa_o^{col}}{\kappa^{col}}\sim\frac{1}{\lambda}$ (here $\lambda$ is the average particle number in a cell), which gives $\kappa_o^{col}\propto\frac{\lambda-1}{\lambda^2\Delta t}\cos\omega\sin\theta$.

    \item Finally, we introduce $k_B$ for dimensional consistency and include a prefactor $\frac{1}{6}$ to match simulation data: $\kappa_o^{\text{col}} = \frac{k_B (\n-1)}{6\n^2\h} \cos\omega\sin\theta$.
\end{enumerate}

\subsubsection{\textbf{The equation of energy conservation with undetermined moments}}
According to \eqr{SRDQ}, the equation of energy conservation for the CSRD fluid is
\begin{equation}
\pd{t}\left( \frac{1}{2}\rho\bk{v^2} \right)+\pd{\a}\left( q_\a-\frac{\h}{2}\pd{t}q_\a \right)=\mathcal{O}(\delta^3)=0,
\end{equation}
where $q_\a=q_\a^{kin}+q_\a^{col}$ and $\pd{t}q_\a$ is
\begin{equation}
\pd{t}q_\a=\pd{t}\left( 2\rho\T u_\a +\frac{1}{2}\rho M_{\b\b\a} \right)+\mathcal{O}(\delta^2).
\end{equation}
The term $\pd{t}M_{\b\b\a}$ involving the third-order moment of velocity is dealt with in the next section and the term $\pd{t}u_\a$ and $\pd{t}\left( \frac{1}{2}\rho\bk{v^2} \right)$ are derived as follows. Substituting $p=\rho\T$ into \eqr{NSeq} gives $\pd{t}u_\a$:
\begin{equation}\label{1stOrderNSeq}
\rho\pd{t}u_\a=-\pd{\a}\left( \rho\T \right)+\mathcal{O}(\delta^2).
\end{equation}
Based on this, the term $\pd{t}\left( \frac{1}{2}\rho\bk{v^2} \right)$ is calculated as
\begin{equation}
\begin{aligned}
    \pd{t}\left( \frac{1}{2}\rho\bk{v^2} \right)&=\pd{t}\left( \rho\T+\frac{1}{2}\rho u^2 \right)\\
                &=\pd{t}\left( \rho\T \right)+\rho u_\a\pd{t}u_\a+\mathcal{O}(\delta^3)\\
                &=\pd{t}\left( \rho\T \right)-\rho u_\a\pd{\a}\left( \rho\T \right)+\mathcal{O}(\delta^3).
\end{aligned}
\end{equation}\par
Therefore, the energy conservation equation becomes
\begin{equation}\label{ThCoEq}
\begin{aligned}
    \pd{t}\left( \rho\T \right)+&\pd{\a}\left( \rho\T u_\a \right)=-\rho\T\pd{\a}u_\a+\pd{\a}\left( \kappa_{\a\b}^{col}\pd{\b} T \right)\\
&+\pd{\a}\left( \rho\T\h\pd{\a}\T \right)+\frac{\h}{4}\pd{\a}\left[ \pd{t}\left( \rho M_{\b\b\a} \right) \right]\\
&-\frac{1}{2}\pd{\a}\left( \rho M_{\b\b\a} \right),
\end{aligned}
\end{equation}
where $\kappa^{col}_{\a\b}=\kappa^{col}\kr{\a\b}+\kappa_o^{col}\ep{\a\b}$. The kinetic thermal conductivity $\kappa^{kin}_{\a\b}$ (contributed due to the streaming step) is contained in the terms of $M_{\b\b\a}$, which will be derived in Section.~\ref{SEC:3rdMoment}.

\subsection{The transport coefficients and hydrodynamic equations of the CSRD fluid}\label{SEC:F}
In this section, the velocity moments $M_{\a\b}$, $M_{\b\b\a}$, and then the viscosities and heat conductivities from the streaming step are separately derived.

\subsubsection{\textbf{The calculation of the velocity moments: general procedure}}

As previously mentioned, $M_{\a\b}$ and $M_{\b\b\a}$ are the velocity moments at position $\bm{r}$, depending on the local equilibrium distribution function $f(\bm{r},\bm{v})$. They can be disturbed slightly under the streaming and collision operations due to the changes of $f(\bm{r},\bm{v})$. The transformations of $f(\bm{r},\bm{v})$ in the streaming and collision steps are denoted by the operators $\hat{\mathcal{S}}$ and $\hat{\mathcal{C}}$ respectively. Therefore, the iteration relation between the distributions at time $t$ and $t+\h$ (denoted by $f_t$ and $f_{t+\h}$) is $f_{t+\h}=\hat{\mathcal{C}}\circ\hat{\mathcal{S}}[f_t]$. According to this iteration equation, the transformations of the moments can be represented as follows.

The zeroth moment, namely the density distribution $\rho$ which is only changed in the streaming step, is
\begin{equation}\label{rhoStr}
\rho^{t+\h}\triangleq\rho^{s}=\IntV \hat{\mathcal{S}}[f_t],
\end{equation}
where the superscript $s$ denotes the distribution after the streaming.

Higher-orders velocity moments before the streaming ($M_{\a\b\ldots}^t$) and after the streaming($M_{\a\b\ldots}^s$) separately read,
\begin{equation}\label{Ms}
\begin{aligned}
    M_{\a\b\ldots}^t&\triangleq\frac{1}{\rho^t}\IntV f_{t}\left( v_\a-u_\a \right)\left( v_\b-u_\b \right)\cdots,\\
    M_{\a\b\ldots}^s&\triangleq\frac{1}{\rho^s}\IntV \hat{\mathcal{S}}[f_{t}]\left( v_\a-u_\a^s \right)\left( v_\b-u_\b^s \right)\cdots,
\end{aligned}
\end{equation}
where $u_\a^s=\frac{1}{\rho^s}\IntV \hat{\mathcal{S}}[f_{t}]v_\a$. The moments after the collision($M_{\a\b\ldots}^{sc}$) are defined by:
\begin{equation}\label{Msc_cplx}
\begin{aligned}
    &M_{\a\b\ldots}^{sc}=M_{\a\b\ldots}^{t+\h}\\
    &\triangleq\frac{1}{\rho^s}\IntV \hat{\mathcal{C}}\circ\hat{\mathcal{S}}[f_{t}]\left( v_\a-u_\a^{sc} \right)\left( v_\b-u_\b^{sc} \right)\cdots,
\end{aligned}
\end{equation}
where ${u_\a^{sc}}=\frac{1}{\rho^s}\int d\bm{v} \hat{\mathcal{C}}\circ\hat{\mathcal{S}}[f_{t}]v_\a$.\par
In a word, the general form of the transform for moments is $M_{\a\b\ldots}^{t+\h}=\hat{\mathcal{F}}[M_{\a\b\ldots}^t]$ and in the steady state the moments do not change with time, so they can be solved by the equation,
\begin{equation}
M_{\a\b\ldots}=\hat{\mathcal{F}}[M_{\a\b\ldots}].
\end{equation}
Thus, what we need to do next is to derive the transform rule $\hat{\mathcal{F}}$, which consists of the streaming and collision contributions.

\qquad\par
\subsubsubsection{The transform of moments in the streaming step}
After the streaming step, $\bm{r}_i(t+\h)=\bm{r}_i(t)+\bm{v}_i(t)\h$,
the distribution becomes $\hat{\mathcal{S}}[f_{t}](\bm{0},\bm{v})=f_{t}(-\bm{v}\h,\bm{v})$, where the position is set to $\bm{r}=\bm{0}$ for simplicity. Based on this relation, we derive the transform of the moments.

The density distribution (the zero-order moment) after the streaming is
\begin{equation}
\rho^{s}=\IntV f_{t}(-\bm{v}\h,\bm{v}),
\end{equation}
which is similar to the integration in \eqr{jy} and can be derived by the variable substitution \eqr{VChange}. Expanding the density distribution about the origin yields,
\begin{equation}\label{rhoChange}
\rho^s=\rho^t-\rho^t\h\pd{\a}u_\a+\mathcal{O}(\delta^2).
\end{equation}
Similarly, other moment after the streaming reads
\begin{equation}
M_{\a\b\ldots}^s\triangleq\frac{1}{\rho^s}\IntV f_{t}(-\bm{v}\h,\bm{v})\left( v_\a-u_\a \right)\left( v_\b-u_\b \right)\cdots,
\end{equation}
and combining with \eqr{rhoChange}, $u_\a,M_{\a\b}$ and $M_{\a\b\g}$ are calculated as,
\begin{equation}\label{uaStr}
\begin{aligned}
    u_\a^s&=u_\a^t-\h\frac{1}{\rho^t}\pd{\a}\left( \rho^t\Tp{t} \right)+\mathcal{O}(\delta^2)\\
    &=u_\a-\h\frac{1}{\rho}\pd{\a}\left( \rho\T \right)+\mathcal{O}(\delta^2),
\end{aligned}
\end{equation}
\begin{equation}\label{MabStr}
M_{\a\b}^s=M_{\a\b}^t-\T\h\left( \pd{\a}u_\b+\pd{\b}u_\a \right)+\mathcal{O}(\delta^2),
\end{equation}
\begin{equation}\label{MabgStr}
\begin{aligned}
    M_{\a\b\g}^s&=M_{\a\b\g}^t-\T\h\left( \kr{\a\b}\kr{\g\t}+\kr{\b\g}\kr{\a\t}+\kr{\g\a}\kr{\b\t} \right)\pd{\t}\T\\
    &+\mathcal{O}(\delta^2).
\end{aligned}
\end{equation}
As a consequence, the traceless part of $M_{\a\b}$ ($M_{\a\b}^\pr$) is,
\begin{equation}\label{Mabp_str}
\begin{aligned}
    M_{\a\b}^{\pr^s}&=M_1^s\bm{\s}_{\a\b}^1+M_3^s\bm{\s}_{\a\b}^3\\
    &=\left( M_1^t-2\T\h e_1 \right)\bm{\s}_{\a\b}^1+\left( M_3^t-2\T\h e_3 \right)\bm{\s}_{\a\b}^3,
\end{aligned}
\end{equation}
with $e_1=\frac{1}{2}\pd{\a}u_\b\bm{\s}_{\a\b}^1=\frac{1}{2}\left( \pd{x}u_y+\pd{y}u_x \right)$ and $e_3=\frac{1}{2}\pd{\a}u_\b\bm{\s}_{\a\b}^3=\frac{1}{2}\left( \pd{x}u_x-\pd{y}u_y \right)$, and the third-order moment $M_{\b\b\a}$ reads,
\begin{equation}\label{Mbba_str}
M_{\b\b\a}^s=M_{\b\b\a}^t-4\T\h\pd{\a}\T.
\end{equation}

\subsubsubsection{The transform of moments in the collision step}\label{SEC::Tran_Col}

The rotation collision in the \eqr{colEq_SM} is an operation acting on all of the particles in the same cell. Thus, one has to consider the state change of each particles when it comes to the transform of the velocity moments. For the cell $\bm{\xi}$, by using the vector $\bm{v}_{\bm{\xi}}=\left( \bm{v}^{(1)},\dots,\bm{v}^{(N)} \right)^\top$, the collision operation \eqr{colEq_SM} can be rewritten as the following format,
\begin{equation}\label{cellCol}
    \bm{v}_{\bm{\xi}}^c=\bm{L}_{\bm{\xi}}\cdot\bm{v}_{\bm{\xi}}\left( t \right),
\end{equation}
where the matrix $\bm{L}_{\bm{\xi}}$ is defined by:
\begin{widetext}
    \begin{equation}\label{Lmatrix}
        \begin{aligned}
            \bm{L}_{\bm{\xi}}  & =
            \begin{bmatrix}
                \bm{R}_{\bm{\xi}}+\frac{1}{N}(\bm{I}-\bm{R}_{\bm{\xi}}) & \frac{1}{N}(\bm{I}-\bm{R}_{\bm{\xi}})                                   & \dots  & \frac{1}{N}(\bm{I}-\bm{R}_{\bm{\xi}})                                   & \\
        
                \frac{1}{N}(\bm{I}-\bm{R}_{\bm{\xi}})                                   & \bm{R}_{\bm{\xi}}+\frac{1}{N}(\bm{I}-\bm{R}_{\bm{\xi}}) & \dots  & \frac{1}{N}(\bm{I}-\bm{R}_{\bm{\xi}})                                   & \\
        
                \vdots                                                                                                                                    & \vdots                                                                                                                                    & \ddots & \vdots                                                                                                                                    & \\
        
                \frac{1}{N}(\bm{I}-\bm{R}_{\bm{\xi}})                                   & \frac{1}{N}(\bm{I}-\bm{R}_{\bm{\xi}})                                   & \dots  & \bm{R}_{\bm{\xi}}+\frac{1}{N}(\bm{I}-\bm{R}_{\bm{\xi}})
            \end{bmatrix}_{2N\times 2N}.
        \end{aligned}
    \end{equation}
\end{widetext}
Here, $\bm{I}$ is the $2\times 2$ unit matrix and $\bm{R}_{\bm{\xi}}=\bm{R}_{\bm{\xi}}\left( \Omega+\theta \right)$. It can be verified that $\bm{L}_{\bm{\xi}}$ is an orthogonal matrix: $\bm{L}_{\bm{\xi}}\bm{L}_{\bm{\xi}}^\top=\bm{I}_{2N\times 2N}$, implying that the phase volume element is conserved in the collision step and so does the CSRD dynamics.

For the sake of convenience, we define some notations as: $\hat{\mathcal{S}}[f_{t}](\bm{r},\bm{v})\triangleq f^s(\bm{r},\bm{v})$,\,$\hat{\mathcal{C}}\circ\hat{\mathcal{S}}[f_{t}](\bm{r},\bm{v})\triangleq f^{sc}(\bm{r},\bm{v})$. We say $f^{sc}(\bm{r},\bm{v})$ is the distribution of the representative particle after the collision. According to the collision operation, this distribution can be written by the distribution before collision:
\begin{equation}\label{fsc}
    f^{sc}(\bm{r},\bm{v})=\ave{f^{s}(\bm{r},\bm{R}^{-1}_{\bm{\xi}}\cdot\left( \bm{v}-\bm{v}_{cm} \right)+\bm{v}_{cm})}.
\end{equation}
The average on the right hand side of \eqr{fsc} can be calculated by introducing the conditional expectation: $\ave{f^{s}(\bm{r},\bm{R}^{-1}\cdot\left( \bm{v}-\bm{v}_{cm} \right)+\bm{v}_{cm})\mid N}$. Here, $N\geqslant 1$ is the particle number in the cell $\bm{\xi}$ containing the representative particle, and $N$ also obeys the distribution of $P(N=q)=e^{-\n}\n^{q-1}/(q-1)!,q\geqslant1$. We label other particles in the cell $\bm{\xi}$ by superscript $(i),\,i=2,\dots,N$. The post-collision conditional probability of particle $i$ (locating in the cell $\bm{\xi}$) having velocity $\bm{v}^{(i)}$ is defined as $p\left( \bm{v}^{(i)} \mid \bm{r}^{(i)}\in A_{\bm{\xi}} \right)$, with $A_{\bm{\xi}}$ the cell area. From the Bayes' formula, we have
\begin{equation}
    \begin{aligned}
        p\left( \bm{v}^{(i)} \mid \bm{r}^{(i)}\in A_{\bm{\xi}} \right)&=\frac{p\left( \bm{v}^{(i)} ; \bm{r}^{(i)}\in A_{\bm{\xi}} \right)}{p\left( \bm{r}^{(i)}\in A_{\bm{\xi}} \right)}\\
        &=\frac{\int_{A_{\bm{\xi}}}d\bm{r}^{(i)}f^{sc}\left( \bm{r}^{(i)},\bm{v}^{(i)} \right)}{\int_{A_{\bm{\xi}}}d\bm{r}^{(i)}\rho^{s}\left( \bm{r}^{(i)} \right)}.
    \end{aligned}
\end{equation}
Using $f^{sc}\left( \bm{r},\bm{v}^{(i)} \right)$ and $\rho^{s}\left( \bm{r} \right)$ to approximate $f^{sc}\left( \bm{r}^{(i)},\bm{v}^{(i)} \right)$ and $\rho^{s}\left( \bm{r}^{(i)} \right)$ in the integration yields,
\begin{equation}
        p\left( \bm{v}^{(i)} \mid \bm{r}^{(i)}\in A_{\bm{\xi}} \right)=\frac{f^{sc}\left( \bm{r},\bm{v}^{(i)} \right)}{\rho^{s}\left( \bm{r} \right)}.
\end{equation}
With the help of $p\left( \bm{v}^{(i)} \mid \bm{r}^{(i)}\in A_{\bm{\xi}} \right)$, \eqr{fsc} can be rewritten as,
\begin{widetext}
    \begin{equation}
        \begin{aligned}
            f^{sc}(\bm{r},\bm{v})&=\ave{f^{s}(\bm{r},\bm{R}^{-1}_{\bm{\xi}}\cdot\left( \bm{v}-\bm{v}_{cm} \right)+\bm{v}_{cm})}\\
            &=\ave{\ave{f^{s}(\bm{r},\bm{R}^{-1}_{\bm{\xi}}\cdot\left( \bm{v}-\bm{v}_{cm} \right)+\bm{v}_{cm})\mid N}}\\
            &=\ave{\int d\bm{v}^{(2)}\cdots d\bm{v}^{(N)}f^{s}(\bm{r},\bm{R}^{-1}_{\bm{\xi}}\cdot\left( \bm{v}-\bm{v}_{cm} \right)+\bm{v}_{cm})p\left( \bm{v}^{(2)},\dots,\bm{v}^{(N)} \mid \bm{r}^{(2)}\in A_{\bm{\xi}},\dots,\bm{r}^{(N)}\in A_{\bm{\xi}} \right)}\\
            &=\ave{\int d\bm{v}^{(2)}\cdots d\bm{v}^{(N)}f^{s}(\bm{r},\bm{R}^{-1}_{\bm{\xi}}\cdot\left( \bm{v}-\bm{v}_{cm} \right)+\bm{v}_{cm})p\left( \bm{v}^{(2)} \mid \bm{r}^{(2)}\in A_{\bm{\xi}} \right)\cdots p\left( \bm{v}^{(N)} \mid \bm{r}^{(N)}\in A_{\bm{\xi}} \right)}\\
            &=\ave{\frac{1}{\left( \rho^s \right)^{N-1}}\int d\bm{v}^{(2)}\cdots d\bm{v}^{(N)}f^{s}(\bm{r},\bm{R}^{-1}_{\bm{\xi}}\cdot\left( \bm{v}-\bm{v}_{cm} \right)+\bm{v}_{cm})f^{sc}\left( \bm{r},\bm{v}^{(2)} \right)\cdots f^{sc}\left( \bm{r},\bm{v}^{(N)} \right)}.
        \end{aligned}
\end{equation}
In the fourth equality, the molecular chaos assumption has been used. Since the distributions $f^{sc}\left( \bm{r},\bm{v}^{(i)} \right)$ in $\ave{\cdot}$ can be expressed by $f^{s}\left( \bm{r},\bm{R}^{-1}_{\bm{\xi}}\cdot\left( \bm{v}^{(i)}-\bm{v}_{cm} \right)+\bm{v}_{cm}\right)$, we have
\begin{equation}
    \begin{aligned}
        f^{sc}(\bm{r},\bm{v})=&\ave{\frac{1}{\left( \rho^s \right)^{N-1}}\int d\bm{v}^{(2)}\cdots d\bm{v}^{(N)}f^{s}(\bm{r},\bm{R}^{-1}_{\bm{\xi}}\cdot\left( \bm{v}-\bm{v}_{cm} \right)+\bm{v}_{cm})\right.\\
        &\left. \cdot f^{s}\left( \bm{r},\bm{R}^{-1}_{\bm{\xi}}\cdot\left( \bm{v}^{(2)}-\bm{v}_{cm} \right)+\bm{v}_{cm} \right)\cdots f^{s}\left( \bm{r},\bm{R}^{-1}_{\bm{\xi}}\cdot\left( \bm{v}^{(N)}-\bm{v}_{cm} \right)+\bm{v}_{cm} \right)}\\
        \triangleq&\ave{\frac{1}{\left( \rho^s \right)^{N-1}}\int d\bm{v}^{(2)}\cdots d\bm{v}^{(N)}\right.\\
        &\left. f^{s}_{\bm{\xi}}\left( \bm{r},\bm{R}^{-1}_{\bm{\xi}}\cdot\left( \bm{v}-\bm{v}_{cm} \right)+\bm{v}_{cm},\bm{R}^{-1}_{\bm{\xi}}\cdot\left( \bm{v}^{(2)}-\bm{v}_{cm} \right)+\bm{v}_{cm},\dots ,\bm{R}^{-1}_{\bm{\xi}}\cdot\left( \bm{v}^{(N)}-\bm{v}_{cm} \right)+\bm{v}_{cm} \right)}\\
        \triangleq&\ave{\frac{1}{\left( \rho^s \right)^{N-1}}\int d\bm{v}^{(2)}\cdots d\bm{v}^{(N)}f^{s}_{\bm{\xi}}\left( \bm{r},\bm{v}^\pr,\bm{v}^{(2)\pr},\dots ,\bm{v}^{(N)\pr} \right)}
    \end{aligned}
\end{equation}
where $f_{\bm{\xi}}^s\left( \bm{r},\bm{v}^\pr,\bm{v}^{(2)\pr},\dots ,\bm{v}^{(N)\pr} \right)=f^s\left( \bm{r},\bm{v}^\pr\right)f^s\left( \bm{r},\bm{v}^{(2)\pr}\right)\cdots f^s\left( \bm{r},\bm{v}^{(N)\pr}\right)$ and $\bm{v}^{(i)\pr}=\bm{R}^{-1}_{\bm{\xi}}\cdot\left( \bm{v}^{(i)}-\bm{v}_{cm} \right)+\bm{v}_{cm}$ are defined. Introducing the vectors $\bm{v}_{\xi}^\pr\triangleq\left( \bm{v}^\pr,\bm{v}^{(2)\pr},\dots ,\bm{v}^{(N)\pr} \right)^\top$ and $\bm{v}_{\xi}\triangleq\left( \bm{v},\bm{v}^{(2)},\dots ,\bm{v}^{(N)} \right)^\top$ and comparing them with Eqs.~\eqref{cellCol},\eqref{Lmatrix}, we can find $\bm{v}_{\xi}^\pr=\bm{L}_{\bm{\xi}}^{-1}\cdot\bm{v}_{\bm{\xi}}$. Therefore, $f^{sc}(\bm{r},\bm{v})$ becomes,
\begin{equation}
    f^{sc}(\bm{r},\bm{v})=\ave{\frac{1}{\left( \rho^s \right)^{N-1}}\int d\bm{v}^{(2)}\cdots d\bm{v}^{(N)}f^{s}_{\bm{\xi}}\left( \bm{r},\bm{L}_{\bm{\xi}}^{-1}\cdot\bm{v}_{\bm{\xi}} \right)}
\end{equation}

Now, we can write down the moments after collision, i.e., \eqr{Msc_cplx}:
\begin{equation}
    \begin{aligned}\label{M_sc}
        M_{\a\b\ldots}^{sc}=&\frac{1}{\rho^s}\IntV f^{sc}\left( \bm{r},\bm{v} \right)\cdot\left( v_\a-u_\a^{sc} \right)\left( v_\b-u_\b^{sc} \right)\cdots=\ave{\frac{1}{\left( \rho^s \right)^{N}}\int d\bm{v_{\bm{\xi}}} f_{\bm{\xi}}^{s}(\bm{r},\bm{L}_{\bm{\xi}}^{-1}\cdot\bm{v}_{\bm{\xi}})\tilde{M}({\bm{v}})},
    \end{aligned}
\end{equation}
where $\tilde{M}(\bm{v})\triangleq\left( v_\a-u_\a^{sc} \right)\left( v_\b-u_\b^{sc} \right)\cdots$. Taking the variable substitution $\bm{v}_{\bm{\xi}}\rightarrow\bm{L}_{\bm{\xi}}^{-1}\bm{v}_{\bm{\xi}}$ for \eqr{M_sc} and together with $\left|\det\left( \bm{L}_{\bm{\xi}} \right)\right|=1$, it can be simplified as
\begin{equation}\label{MomentCol}
    \begin{aligned}
        M_{\a\b\ldots}^{sc}&=\ave{\frac{1}{\left( \rho^s \right)^{N}}\int d\bm{v}_{\bm{\xi}}f_{\bm{\xi}}^s(\bm{r},\bm{v}_{\bm{\xi}})\tilde{M}\left( \frac{N-1}{N}\left[ \bm{R_\xi}\left( \bm{v}-\hat{\bm{v}} \right) +\hat{\bm{v}} \right]+\frac{1}{N}\bm{v} \right)}\\
                &=\ave{\bk{\tilde{M}\left( \frac{N-1}{N}\left[ \bm{R_\xi}\cdot\left( \bm{v}-\hat{\bm{v}} \right) +\hat{\bm{v}} \right]+\frac{1}{N}\bm{v} \right)}^s}\\
                &=\ave{\bk{\tilde{M}\left( \bm{v}^c \right)}^s},
    \end{aligned}
\end{equation}
where the decomposition $\bm{v}_{cm}=\frac{1}{N}\bm{v}+\frac{N-1}{N}\hat{\bm{v}}$ has been used and $\bm{v}^c$ is (also see \eqr{colDetailEq})
\begin{equation}\label{colDetailEq2}
\begin{aligned}
    v_\a^c&=\frac{N-1}{N}\left[ \left( \cos\a+\frac{1}{N-1} \right)v_\a-\sin\a v_\a^\ast+\left( 1-\cos\a \right)\hat{v}_\a+\sin\a\hat{v}_\a^\ast\right]\triangleq c_1v_\a-s_1v_\a^\ast+c_2\hat{v}_\a+s_1\hat{v}_\a^\ast.
\end{aligned}
\end{equation}
Here, $c_1$ and $c_2$ satisfy the relation $c_1+c_2=1$.

\eqr{MomentCol} is the general form of the moments transform in the collision step. The transforms of the second- and third-order moments are derived as follows. For convenience, we define the notations $\bbkk{\cdot}\triangleq\ave{\bk{\cdot}^s}$ in the remaining part.

For the second-order moment, setting $\tilde{M}(\bm{v}^c)=\left( v_\a^c-u_\a^s \right)\left( v_\b^c-u_\b^s \right)$ in \eqr{MomentCol} gives
\begin{equation}
\begin{aligned}
     M_{\a\b}^{sc}&=\bbkk{\left( v_\a^c-u_\a^s \right)\left( v_\b^c-u_\b^s \right)}=\bbkk{v_\a^c v_\b^c}-\bbkk{v_\a^c}u_\b^s-\bbkk{v_\b^c}u_\a^s+u_\a^s u_\b^s,
\end{aligned}
\end{equation}
where $\bbkk{v_\a^c}$, the average velocity, is
\begin{equation}\label{Mc1}
\begin{aligned}
    \bbkk{v_\a^c}&=\ave{c_1}u_\a^s-\ave{s_1}{u_\a^s}^\ast+\ave{c_2}u_\a^s+\ave{s_1}{u_\a^s}^\ast=\ave{c_1+c_2}u_\a^s=u_\a^s.
\end{aligned}
\end{equation}
Thus, the second moment is
\begin{equation}\label{Mab0}
M_{\a\b}^{sc}=\bbkk{v_\a^c v_\b^c}-u_\a^s u_\b^s=\bbkk{v_\a^c v_\b^c}+\mathcal{O}(\delta^2).
\end{equation}
Using the molecular chaos assumption and \eqr{colDetailEq2}, $\bbkk{v_\a^c v_\b^c}$ becomes
    \begin{equation}\label{vavb}
        \begin{aligned}
            \bbkk{v_\a^c v_\b^c}=&\,M_{\a\b}^s\ave{c_1^2-s_1^2+\frac{1}{N-1}(1-c_1)^2-\frac{1}{N-1}s_1^2}+\kr{\a\b}M_{\g\g}^s\ave{s_1^2\left( 1+\frac{1}{N-1} \right)}\\
            &+\left( \ep{\b\g}M_{\a\g}^s +\ep{\a\g}M_{\g\b}^s \right)\ave{\frac{1}{N-1}(1-c_1)s_1-c_1s_1}+\mathcal{O}(\delta^2)\\
            =&\,M_{\a\b}^s\ave{\frac{N-1}{N}\cos2\a+\frac{1}{N}}-\ep{\b\g}M_{\a\g}^s\ave{\frac{N-1}{2N}\sin2\a}-\ep{\a\g}M_{\g\b}^s\ave{\frac{N-1}{2N}\sin2\a}\\
               &+\kr{\a\b}M_{\g\g}^s\ave{\frac{N-1}{2N}(1-\cos2\a)}+\mathcal{O}(\delta^2).
        \end{aligned}
    \end{equation}
Then, the transform of the second-order moment in collision is obtained as
\begin{equation}\label{MabCol}
\begin{aligned}
    M_{\a\b}^{sc}=&\,\left( 1-2c \right)M_{\a\b}^s-s\left( \ep{\b\g}M_{\a\g}^s+\ep{\a\g}M_{\g\b}^s \right)+c\kr{\a\b}M_{\g\g}^s+\mathcal{O}(\delta^2),
\end{aligned}
\end{equation}
where the coefficients $c$ and $s$ are
\begin{equation}
\begin{aligned}
    c&=\ave{\frac{N-1}{2N}\left( 1-\cos2\a \right)}=\frac{\n-1+e^{-\n}}{2\n}\left( 1-\cos2\theta\cos2\omega \right),
\end{aligned}
\end{equation}
\begin{equation}
    s=\ave{\frac{N-1}{2N}\sin2\a}=\frac{\n-1+e^{-\n}}{2\n}\sin2\theta\cos2\omega.
\end{equation}

Similar procedure can be performed to obtain the transform of the third-order moment. Taking $\tilde{M}(\bm{v}^c)=\left( v_\a^c-u_\a^s \right)\left( v_\b^c-u_\b^s \right)\left( v_\g^c-u_\g^s \right)$ \eqr{MomentCol} and using \eqr{colDetailEq2} gives $M_{\a\b\g}^{sc}$,
\begin{equation}\label{MabgCol}
\begin{aligned}
    M_{\a\b\g}^{sc}=&\,A_1M_{\a\b\g}^s-A_2\left( \ep{\g\t}M_{\a\b\t}^s+\ep{\b\t}M_{\a\t\g}^s+\ep{\a\t}M_{\t\b\g}^s \right)\\
                &-A_3\left( \ep{\g\t}\kr{\a\b}+\ep{\b\t}\kr{\a\g}+\ep{\a\t}\kr{\b\g} \right)M_{\t\nu\nu}^s\\
                &+A_4\left( \kr{\b\g}M_{\a\t\t}^s +\kr{\a\g}M_{\b\t\t}^s+\kr{\a\b}M_{\g\t\t}^s\right)+\mathcal{O}(\delta^2),
\end{aligned}
\end{equation}
where
\begin{equation}\label{coefficientA}
    \begin{aligned}
        A_1&=\ave{\frac{1}{N^2}\left\{ 1+\left( N-1 \right)\left[ 2\left( N-2 \right)\cos\a\cos2\a+3\cos2\a-\left( N-2 \right)\cos\a \right] \right\}},\\
        A_2&=\ave{\frac{N-1}{3N^2}\left[ 2\left( N-2 \right)\sin\a\cos2\a+3\sin2\a+\left( N-2 \right)\sin\a \right]},\\
        A_3&=\ave{\frac{1}{6N^2}\left( N-1 \right)\left( N-2 \right)\sin\a\left( 1-\cos2\a \right)},\\
        A_4&=\ave{\frac{N-1}{2N^2}\left[ 1+\left( N-2 \right)\cos\a \right]\left( 1-\cos2\a \right)}.
    \end{aligned}
\end{equation}

Typically, the transforms of moments that need to be calculated ($M_{\a\b}^\pr$ in \eqr{Stress0} and $M_{\b\b\a}$ in \eqr{ThCoEq}) are
\begin{equation}\label{Mabp_col}
\begin{aligned}
    {M_{\a\b}^\pr}^{sc}=&\,M_1^{sc}\bm{\sigma}^1_{\a\b}+M_3^{sc}\bm{\sigma}^3_{\a\b}=\left[ \left( 1-2c \right)M_1^s + 2sM_3^s \right]\bm{\sigma}^1_{\a\b}+\left[ \left( 1-2c \right)M_3^s - 2sM_1^s \right]\bm{\sigma}^3_{\a\b},
\end{aligned}
\end{equation}
and
\begin{equation}\label{Mbba_col}
\begin{aligned}
    M_{\b\b\a}^{sc}&=\left[ \left( A_1+4A_4 \right)\kr{\a\g}-\left( A_2+4A_3 \right)\ep{\a\g} \right]M_{\b\b\g}^s\triangleq\left( A_e\kr{\a\g}-A_o\ep{\a\g} \right)M_{\b\b\g}^s.
\end{aligned}
\end{equation}
\end{widetext}

\subsubsection{\textbf{The calculation of the velocity moments: derivations}}
\subsubsubsection{The derivation of the second-order moment}\label{SEC:2ndMoment}
The complete transform of $M_{\a\b}^\pr$ is derived through the results \eqr{Mabp_str} and \eqr{Mabp_col}:
\begin{equation}
\begin{aligned}
    &M_{\a\b}^{\pr^{t+\h}}\\
    &=[ \left( 1-2c \right)\left( M_1^t-2\T\h e_1 \right)+2s\left( M_3^t-2\T\h e_3 \right) ]\bm{\s}_{\a\b}^1\\
    &+[ \left( 1-2c \right)\left( M_3^t-2\T\h e_3 \right)-2s\left( M_1^t-2\T\h e_1 \right) ]\bm{\s}_{\a\b}^3.
\end{aligned}
\end{equation}
Setting $M_{\a\b}^{\pr^{t+\h}}=M_{\a\b}^{\pr^t}$ gives a equation for the steady-state $M'_{\a\b}$:
\begin{equation}\label{EqOfMabPri}
\begin{bmatrix}
    c & -s \\
    s & c
\end{bmatrix}
\begin{bmatrix}
    M_1\\
    M_3
\end{bmatrix}
=-\T\h
\begin{bmatrix}
    \left( 1-2c \right)e_1+2s e_3\\
    \left( 1-2c \right)e_3-2s e_1
\end{bmatrix},
\end{equation}
whose solution is:
\begin{equation}\label{SolvedMabPri}
\begin{aligned}
    M_1&=-\T\h\left[ \left( p_e-2 \right)e_1 + p_o e_3 \right],\\
    M_3&=\T\h\left[ p_o e_1 - \left( p_e-2 \right)e_3 \right],
\end{aligned}
\end{equation}
where
\begin{equation}
p_e=\frac{c}{s^2+c^2},\qquad p_o=\frac{s}{s^2+c^2}.
\end{equation}

\subsubsubsection{The derivation of the third-order moment}\label{SEC:3rdMoment}
The iteration relation of the third-order moment $M_{\b\b\a}$ between $t$ and $t+\h$ is given by \eqr{Mbba_str} and \eqr{Mbba_col}:
\begin{equation}\label{MbbaIter}
M_{\b\b\a}^{t+\h}=\left( A_e\kr{\a\g}-A_o\ep{\a\g} \right)\left(M_{\b\b\g}^t-4\T\h\pd{\g}\T\right).
\end{equation}
Thus, in the steady state, $M_{\b\b\a}$ can be solved by the following equation:
\begin{widetext}
    \begin{equation}\label{EqOfMbba}
    \begin{aligned}
        &\begin{bmatrix}
            A_e-1 & A_o \\
            -A_o & A_e-1
        \end{bmatrix}
        \begin{bmatrix}
            M_{\b\b y} \\
            M_{\b\b x}
        \end{bmatrix}
        =4\T\h
        \begin{bmatrix}
            A_e\pd{y}\T+A_o\pd{x}\T \\
            -A_o\pd{y}\T+A_o\pd{x}\T
        \end{bmatrix}.
    \end{aligned}
    \end{equation}
\end{widetext}
The solution of \eqr{EqOfMbba} is
\begin{equation}\label{SolvedMbba0}
\begin{aligned}
    M_{\b\b y}&=4\T\h\left[ \left( 1+\vp_e \right)\pd{y}\T-\vp_o \pd{x}\T \right],\\
    M_{\b\b x}&=4\T\h\left[ \left( 1+\vp_e \right)\pd{x}\T+\vp_o \pd{y}\T \right],
\end{aligned}
\end{equation}
whose equivalent form is
\begin{equation}\label{SolvedMbba}
M_{\b\b\a}=4\T\h\left[ \left( 1+\vp_e \right)\pd{\a}\T+\vp_o \pd{\a}^\ast\T \right],
\end{equation}
where the definitions of $\vp_e$ and $\vp_o$ are
\begin{equation}\label{vpeAndvpo}
\vp_e=\frac{A_e-1}{\left( A_e-1 \right)^2+A_o^2},\qquad\vp_o=\frac{A_o}{\left( A_e-1 \right)^2+A_o^2}.
\end{equation}

\subsubsection{\textbf{The constitutive relation and the Navier-Stokes equation}}
Now, we can obtain the stress tensor by substituting the expression of $M_{\a\b}$ (\eqr{SolvedMabPri}) into \eqr{Stress0}:
\begin{equation}\label{Stress1}
\begin{aligned}
    \sigma_{\a\b}=&\,-p\kr{\a\b}+\eta^{kin}\left( \pd{\b}u_\a+\pd{\a}u_\b-\kr{\a\b}\pd{\g}u_\g \right)\\
                    &+\eta_o^{kin}\left( \pd{\b}u_\a^\ast+\pd{\b}^\ast u_\a \right)+\eta^{col}\pd{\b}u_\a+\eta_o^{col}\pd{\b}u_\a^\ast,
\end{aligned}
\end{equation}
where the normal and odd viscosities from the kinetic contribution are separately,
\begin{widetext}
    \begin{equation}
        \eta^{kin}=\frac{1}{2}\rho \T\h\left( e_e-1 \right)=nk_BT\h\left[\frac{\n}{\n-1+e^{-\n}}\frac{1-\cos2\theta\cos2\omega}{1+\cos2\omega(\cos2\omega-2\cos2\theta)}-\frac{1}{2}\right],
        \end{equation}
        \begin{equation}
        \eta_o^{kin}=-\frac{1}{2}\rho\T\h e_o=-nk_BT\h\frac{\n}{\n-1+e^{-\n}}\frac{\sin2\theta\cos2\omega}{1+\cos2\omega(\cos2\omega-2\cos2\theta)}.
        \end{equation}
\end{widetext}
Following the notations in \cite{Vitelli_Fluhydro}, the standard form of constitutive relation for the CSRD fluid is,
\begin{equation}\label{ConEq}
    \begin{aligned}
        \sigma_{\alpha\beta}=&-p\kr{\a\b}+\left[\eta\left(\delta_{\alpha\mu}\delta_{\beta\nu}+\delta_{\alpha\nu}\delta_{\beta\mu}-\delta_{\alpha\beta}\delta_{\mu\nu}\right)\right.\\
                            &+\eta_R\left(\delta_{\alpha\mu}\delta_{\beta\nu}-\delta_{\alpha\nu}\delta_{\beta\mu}\right)+\zeta\delta_{\alpha\beta}\delta_{\mu\nu}\\
                            &+\eta_o\left(\varepsilon_{\alpha\mu}\delta_{\beta\nu}+\varepsilon_{\beta\nu}\delta_{\alpha\mu}\right)\\
                            &\left.-\eta_A\ep{\a\b}\kr{\mu\nu}-\eta_B\kr{\a\b}\ep{\mu\nu}\right]\partial_\nu u_\mu,
    \end{aligned}
\end{equation}
where $\eta=\eta^{kin}+\frac{1}{2}\eta^{col}$ is the shear viscosity, $\zeta=\frac{1}{2}\eta^{col}$ the bulk viscosity, $\eta_R=\frac{1}{2}\eta^{col}$ the rotation-rotation viscosity, $\eta_o=\eta^{kin}_o+\frac{1}{2}\eta^{col}_o$ the odd viscosity, $\eta_A=-\frac{1}{2}\eta^{col}_o$ the compression-rotation viscosity, and $\eta_B=\frac{1}{2}\eta^{col}_o$ the rotation-compression viscosity.

Finally, inserting the constitutive relation into \eqr{NSeq}, the Navier-Stokes equation for the CSRD fluid becomes,
\begin{equation}
\pd{t}\left( \rho \bm{u} \right) + \nabla\cdot\left( \rho\bm{u}\bm{u} \right)=-\nabla p+\hat{\eta}\nabla^2\bm{u}+\hat{\eta}_o \bm{\varepsilon}\cdot\nabla^2\bm{u},
\end{equation}
where $\hat{\eta}=\eta^{kin}+\eta^{col}$ and $\hat{\eta}_o=\eta^{kin}_o+\eta^{col}_o$.

\subsubsection{\textbf{The equation of heat conduction and the thermal conductivities}}
The equation of heat conduction derived in the previous section is
\begin{equation}
\begin{aligned}
    \pd{t}\left( \rho\T \right)&+\pd{\a}\left( \rho\T u_\a \right)=-\rho\T\pd{\a}u_\a+\pd{\a}\left( \kappa_{\a\b}^{col}\pd{\b} T \right)\\
&+\pd{\a}\left( \rho\T\h\pd{\a}\T \right)+\frac{\h}{4}\pd{\a}\left[ \pd{t}\left( \rho M_{\b\b\a} \right) \right]\\
&-\frac{1}{2}\pd{\a}\left( \rho M_{\b\b\a} \right).
\end{aligned}
\end{equation}
From this equation, the heat flux contributed by the streaming step reads $q^{kin}_\a\triangleq \rho\T\h\pd{\a}\T+\frac{\h}{4}\pd{t}\left( \rho M_{\b\b\a} \right)-\frac{1}{2}\rho M_{\b\b\a}$. Substituting $M_{\b\b\a}$(\eqr{SolvedMbba}) into the kinetic heat flux gives
\begin{equation}
\begin{aligned}
    q^{kin}_\a=&\,-\rho\T\h\pd{\a}\T-\frac{\h}{4}\pd{t}\left( \rho M_{\b\b\a} \right)+\frac{1}{2}\rho M_{\b\b\a}\\
    =&\,\rho\T\h\left[ \left( 1+2\vp_e \right)\pd{\a}\T+2\vp_o \pd{\a}^\ast\T \right]+\mathcal{O}(\delta^2)\\
    \triangleq&\, -\frac{m}{k_B}\kappa^{kin}\pd{\a}\T-\frac{m}{k_B}\kappa_o^{kin}\pd{\a}^\ast\T\\
    =&\,-\kappa^{kin}\pd{\a} T-\kappa_o^{kin}\pd{\a}^\ast T,
\end{aligned}
\end{equation}
where $\kappa^{kin}$ and $\kappa_o^{kin}$ are, respectively, the normal (even) and odd thermal conductivities contributed by the streaming step:
\begin{equation}
\begin{aligned}
    \kappa^{kin}&=\frac{k_B}{m}\left[-2\left( 1+\vp_e \right)-\vp_e \ovl{b_1}+\vp_o \ovl{b_2} \right]\rho\T\h,\\
    \kappa_o^{kin}&=\frac{k_B}{m}\left( -2\vp_o-\vp_e\ovl{b_2}-\vp_o\ovl{b_1}\right)\rho\T\h.
\end{aligned}
\end{equation}
For the large $\n$ limit, replace $N$ by $\n$ in $A_e$ and $A_o$ (see Eqs.~\eqref{coefficientA}, \eqref{Mbba_col} and \eqref{vpeAndvpo}), then $\kappa^{kin}$ and $\kappa_o^{kin}$ read:
\begin{widetext}
    \begin{equation}
        \begin{aligned}
            &\kappa^{kin}=n\frac{k_B^2 T}{m}\h\left\{ \frac{\parbox{11cm}{\centering$\n\left[ \left( \n+2\cos\omega \right)\sin^2\frac{\omega}{2}+\left( \left( \n-2 \right)\cos\omega+4\cos2\omega\cos^2\frac{\theta}{2} \right)\sin^2\frac{\theta}{2} \right]$}}{\parbox{11cm}{\centering$\left( \n-1 \right)\left\{ \left( \n+2\cos\omega \right)^2\sin^4\frac{\omega}{2}\right.$\\$\left.+\left[ \left( \n-2 \right)\left( \n-1-\cos2\omega \right)\cos\omega+4\left( \n-1 \right)\cos2\omega\cos^2\frac{\theta}{2} \right]\sin^2\frac{\theta}{2} \right\}$}} -1 \right\},\\
            &\kappa_o^{kin}=-n\frac{k_B^2 T}{m}\h\frac{\parbox{11cm}{\centering$\n^2\left[ \left( \n-2 \right)\cos\omega+2\cos2\omega\cos\theta \right]\sin\theta$}}{\parbox{11cm}{\centering$2\left( \n-1 \right)\left\{ \left( \n+2\cos\omega \right)^2\sin^4\frac{\omega}{2}\right.$\\$\left.+\left[ \left( \n-2 \right)\left( \n-1-\cos2\omega \right)\cos\omega+4\left( \n-1 \right)\cos2\omega\cos^2\frac{\theta}{2} \right]\sin^2\frac{\theta}{2} \right\}$}}.
        \end{aligned}
        \end{equation}
\end{widetext}
Finally, the equation of heat conduction for the CSRD fluid is,
\begin{equation}\label{HCeq1}
\pd{t}\left( \rho\T \right)+\pd{\a}\left( \rho\T u_\a \right)=-\rho\T\pd{\a}u_\a+\pd{\a}\left( \kappa_{\a\b}\pd{\b}T \right),
\end{equation}
or equivalently,
\begin{equation}\label{HCeq2}
    \frac{k_B}{m}\left[ \pd{t}\left( \rho T \right)+\pd{\a}\left( \rho T u_\a \right) \right]=-p\pd{\a}u_\a+\pd{\a}\left( \kappa_{\a\b}\pd{\b}T \right),
\end{equation}
with
\begin{equation}
\kappa_{\a\b}=\left( \kappa^{kin}+\kappa^{col} \right)\kr{\a\b}+\left( \kappa_o^{kin}+\kappa_o^{col} \right)\ep{\a\b}.
\end{equation}
Note that, in general, both the pressure and viscous stress can contribute to the mechanical work. However, in the above heat conduction equation, there exists only the pressure contribution ($-p\pd{\a}u_\a$), while the term related to the mechanical work from the viscous stress (i.e., $\sigma_{\a\b}+p\kr{\a\b}$) is absent because of the limitation of the kinetic theory used here. By considering the conservation of energy, we can correct the heat conduction equation (Eqs.~\eqref{HCeq2}) by adding the viscous stress term phenomenologically. The resulted heat conduction equation becomes,
\begin{equation}
    \frac{k_B}{m}\left[ \pd{t}\left( \rho T \right)+\pd{\a}\left( \rho T u_\a \right) \right]=\sigma_{\a\b}\pd{\b}u_\a+\pd{\a}\left( \kappa_{\a\b}\pd{\b}T \right).
\end{equation}

\renewcommand\thefigure{C\arabic{figure}}  
\renewcommand\thetable{C\arabic{table}}  
\renewcommand{\theequation}{C\arabic{equation}}
\setcounter{equation}{0}
\setcounter{figure}{0} 
\setcounter{table}{0}
\section{Theoretical derivation of self-diffusion coefficient of the CSRD fluid}\label{APPENDIX::C}
The self-diffusion behavior of the CSRD fluid is studied in this section. Here, the CSRD particles are distinguished by labelling some particles as $A$ and other particles as $B$, but their dynamics are completely identical. Thus, the CSRD fluid can be regarded as a binary mixture. Near equilibrium, the evolution of the density difference $\Delta\rho=\rho_A-\rho_B$ between species $A$ and $B$ is described by the self-diffusion equation,
\begin{equation}
\pd{t}\Delta\rho=\nabla\cdot\bm{J}^D,
\end{equation}
where $\bm{J}^D=\bm{J}^{(m_A)}-\bm{J}^{(m_B)}$ is the self-diffusion flux. With the discrete self-diffusion flux $j_\a^D=j_\a^{(m_A)}-j_\a^{(m_B)}$ ($j_\a^{(m_I)},\,I\in\left\{ A,B \right\}$ is the discrete mass flux of species $I$), according to \eqr{SRDQ} the continuity equation becomes,
\begin{equation}\label{SDIF0}
\pd{t}\Delta\rho+\pd{\a}\left( j_\a^D-\frac{\h}{2}\pd{t}j_\a^D \right)=\mathcal{O}(\delta^3).
\end{equation}

In the above continuity equation, the density of species $I$ is expressed as $\rho_I(\bm{r})=\IntV f_I(\bm{r},\bm{v})$, with $f_I(\bm{r},\bm{v})$ the single-particle distribution of species $I$. Accordingly, the average of quantity $X$ for species $I$ is written as,
\begin{equation}
\bk{X_I(\bm{r})}\triangleq\frac{1}{\rho_I(\bm{r})}\IntV Xf_I(\bm{r},\bm{v}).
\end{equation}
To proceed, $f_I(\bm{r},\bm{v})$ is considered as the local equilibrium one, being of the same functional form as (\eqr{localEqdistribution}),
\begin{equation}
f_I(\bm{r},\bm{v})=\frac{\rho_I(\bm{r})}{\T_I(\bm{r})}g\left( \frac{\bm{v}-\bm{u}_I(\bm{r})}{\sqrt{\T_I(\bm{r})}} \right),
\end{equation}
where $\bm{u}_I$ and $\T_I$ are the mean velocity and temperature of the particles of species $I$, respectively. Thus, according to \eqr{jm}, the discrete mass flux reads
\begin{equation}
\begin{aligned}
    j_\a^{(m_I)}=&\,\rho_Iu_{I,\a}-\frac{1}{2}\h\pd{\a}{\left( \rho_I\T_I \right)}-\frac{1}{2}\h\pd{\b}{\left( \rho u_{I,\b} u_{I,\a} \right)}\\
    &+\mathcal{O}(\delta^2).
\end{aligned}
\end{equation}
Hence, the self-diffusion flux $j_\a^D$ is derived as,
\begin{equation}\label{JD}
\begin{aligned}
    j_\a^D=&\,j_\a^{(m_A)}-j_\a^{(m_B)}\\
          =&\,\rho_A u_{A,\a}-\rho_B u_{B,\a} -\frac{1}{2}\h\pd{\a}\left( \rho_A\T_A-\rho_B\T_B \right)\\
           &+\mathcal{O}(\delta^2)\\
          =&\,\rho_A u_{A,\a}-\rho_B u_{B,\a} -\frac{1}{2}\h\pd{\a}\left( \Delta\rho\T \right)+\mathcal{O}(\delta^2).
\end{aligned}
\end{equation}
Here, we assume that the system is in isothermal and mechanical equilibrium conditions and there is no flow in the system. Thus these conditions imply $\T_A=\T_B=\T=const.$, $p=const.$, and $u_\a=0$.

The first term of \eqr{JD}, i.e., $\mathfrak{m}_\a\triangleq\rho_A u_{A,\a}-\rho_B u_{B,\a}$, can be calculated by the same method, as has been used to derive the velocity moments. In terms of \eqr{rhoChange} and \eqr{uaStr}, the transform of $\mathfrak{m}_\a$ in the streaming step is
\begin{equation}\label{m_str}
\mathfrak{m}_\a^s=\mathfrak{m}_\a-\h\pd{\a}\left( \Delta\rho\T \right)+\mathcal{O}(\delta^2).
\end{equation}
Then we turn to derive its transformation in the collision step. After collision $\mfr_\a^s$ becomes $\mfr_\a^{sc}$ which is expressed by
\begin{equation}\label{m_col_1}
    \mfr_\a^{sc}=\int\mathrm{d}\bm{v}\left[ f_A^{sc}\left( \bm{r},\bm{v} \right) - f_B^{sc}\left( \bm{r},\bm{v} \right) \right]v_\a.
\end{equation}
We first focus on the first term of Eq.~\eqref{m_col_1}. Similar to the derivation of $f^{sc}$ in Section~\ref{SEC::Tran_Col}, the distribution function for species $A$ after collision can be derived as
\begin{equation}
    \begin{aligned}
        f_A^{sc}&=\ave{f_A^s\left( \bm{r}, \bm{R}_{\bm{\xi}}^{-1}\cdot\left( \bm{v} - \bm{v}_{cm} \right)+\bm{v}_{cm} \right)}\\
                &=\ave{\frac{1}{\left( \rho^s \right)^{N-1}}\int\mathrm{d}\bm{v}^{(2)}\cdots\mathrm{d}\bm{v}^{(N)}f^s_{A|\cdots}\left( \bm{r},\bm{L}_{\bm{\xi}}^{-1}\cdot\bm{v}_{\bm{\xi}} \right)},
    \end{aligned}
\end{equation}
where
\begin{equation}
    f^s_{A|\cdots}\left( \bm{r},\bm{v}_{\bm{\xi}} \right)\triangleq f_A^s\left( \bm{r}, \bm{v} \right)f^s\left( \bm{r}, \bm{v}^{(2)} \right)\cdots f^s\left( \bm{r}, \bm{v}^{(N)} \right).
\end{equation}
Substituting this into $\int\mathrm{d}\bm{v}f_A^{sc}\left( \bm{r},\bm{v} \right)v_\a$ yields
\begin{equation}
    \begin{aligned}
        \int\mathrm{d}\bm{v}f_A^{sc}\left( \bm{r},\bm{v} \right)v_\a=\rho_A^s u_{A,\a}^s - \rho_A^s\ave{\mathcal{C}_{\a\b}}\left( u_{A,\a}^s - u_\a^s \right),
    \end{aligned}
\end{equation}
where $\mathcal{C}_{\a\b}$ is defined as
\begin{equation}
    \mathcal{C}_{\a\b}=\frac{N-1}{N}\left( \kr{\a\b}-R_{\a\b} \right).
\end{equation}
By applying the same derivation, we obtain the second term in Eq.~\eqref{m_col_1}:
\begin{equation}
    \int\mathrm{d}\bm{v}f_B^{sc}\left( \bm{r},\bm{v} \right)v_\a=\rho_B^s u_{B,\a}^s - \rho_B^s\ave{\mathcal{C}_{\a\b}}\left( u_{B,\a}^s - u_\a^s \right).
\end{equation}
Therefore, the result of $\mfr_\a^{sc}$ is
\begin{equation}\label{m_col}
    \begin{aligned}
        \mfr_\a^{sc} &= \left( \kr{\a\b} - \ave{\mathcal{C}_{\a\b}} \right)\mfr_\b^s + \Delta\rho^s\ave{\mathcal{C}_{\a\b}}u_{\b}^s\\
        &=\left( \kr{\a\b} - \ave{\mathcal{C}_{\a\b}} \right)\mfr_\b^s.
    \end{aligned}
\end{equation}
The total transformation of $\mfr$ is then derived from Eqs.~\eqref{m_str} and Eqs.~\eqref{m_col}：
\begin{equation}
    \mfr_\a^{sc} = \left( \kr{\a\b} - \ave{\mathcal{C}_{\a\b}} \right)\left[ \mfr_\b - \h\pd{\b}\left( \Delta\rho\T \right) \right].
\end{equation}
Setting $\mfr_\a^{sc}=\mfr_\a^{t}=\mfr_\a$ in this equation, we obtain the solution for the stationary value of $\mfr_\a$:
\begin{equation}
    \mfr_\a=\left( \kr{\a\b} - \ave{\mathcal{C}_{\a\b}}^{-1} \right)\h\pd{\b}\left( \Delta\rho\T \right).
\end{equation}

Thus, the self-diffusion flux \eqr{JD} becomes
\begin{equation}
    j_\a^D=\left( \frac{1}{2}\kr{\a\b} - \ave{\mathcal{C}_{\a\b}}^{-1} \right)\h\pd{\b}\left( \Delta\rho\T \right).
\end{equation}
Substituting this into the conservation equation \eqr{SDIF0}, the self-diffusion equation is obtained,
\begin{equation}
\pd{t}\Delta\rho=\pd{\a}\pd{\b}\left( D_{\a\b}\Delta\rho \right),
\end{equation}
where $D_{\a\b}$ is the self-diffusion tensor:
\begin{equation}
D_{\a\b}=-\T\h\left( \frac{1}{2}\kr{\a\b} - \ave{\mathcal{C}_{\a\b}}^{-1} \right)\triangleq D\kr{\a\b}+D_o\ep{\a\b}.
\end{equation}
Here, the subscript $o$ represents the odd diffusion coefficients, and the specific expressions of the diffusion coefficients are
\begin{widetext}
    \begin{equation}
        \begin{aligned}
            D=&\frac{k_BT\h}{2m}\left[ \frac{2\n\left( 1-\cos\omega\cos\theta \right)}{\left( \n-1+e^{-\n} \right)\left( 1-2\cos\omega\cos\theta+\cos^2\omega \right)}-1 \right],\\
            D_o=& -\frac{\n k_BT\h\cos\omega\sin\theta}{m\left( \n-1+e^{-\n} \right)\left( 1-2\cos\omega\cos\theta+\cos^2\omega \right)}. 
        \end{aligned}
        \end{equation}
\end{widetext}
Finally, the self-diffusion equation reads
\begin{equation}
\pd{t}\Delta\rho=D\nabla^2\Delta\rho.
\end{equation}

\end{document}